\documentclass{JFM-FLM_Au}
\usepackage{setspace}
\usepackage{graphicx}
\usepackage{hyperref}
\usepackage{subcaption}
\usepackage{pdflscape}
\usepackage{amsmath}

\lefttitle{Zhengqin Shu and Chunxiao Xu}
\righttitle{Journal of Fluid Mechanics}

\title{Mean velocity profiles and total shear stress profiles in adverse-pressure-gradient turbulent boundary layers considering history effect}

\author{Zhengqin Shu\aff{1}, \and Chunxiao Xu\aff{2}}

\affiliation{\aff{1} TEEP, Xingjian College, Tsinghua University, Beijing 100084, China
\aff{2} AML, Department of Engineering Mechanics, Tsinghua University, Beijing 100084, China}
\corresau{Chunxiao Xu, \email{xucx@mail.tsinghua.edu.cn}}

\date{\today}

\begin{document}

\maketitle

\begin{abstract}

    This study focuses on developing a predictive model for mean velocity profiles and total shear stress profiles in turbulent boundary layers subjected to adverse pressure gradients, especially with history effects. 
    A new scaling using friction velocity modified by Clauser pressure gradient parameter is introduced to restore streamwise self-similarity.
    Furthermore, an estimation-correction model is developed, explicitly incorporating a streamwise derivative of pressure gradient, which effectively captures history effect beyond the reach of Reynolds-averaged Navier-Stokes equations.
    With the help of the model, the total shear stress is decomposed into four parts, representing respectively the Reynolds number effects, equilibrium pressure gradient effects, the coupling between free-stream velocity and pressure gradient, and local non-equilibrium pressure gradient effects. The latter two are considered first-order history effects, and can account for up to approximately half of the total stress.
    Validation against multiple DNS/LES datasets across a wide range of pressure gradients and Reynolds numbers demonstrates the model's accuracy in predicting both mean velocity profiles and total shear stress profiles. 
\end{abstract}

\section{Introduction}\label{sec:Introduction}

Turbulent boundary layers (TBLs) are essential to understanding many natural and engineering flow systems. Since L. Prandtl's boundary layer theory, TBLs have formed the basis for drag prediction, heat transfer control, and turbulence modeling. 
In real-world natural or engineering environments, TBLs are commonly subjected to streamwise pressure gradients (PGs), including both favorable pressure gradient (FPG) and adverse pressure gradient (APG) cases, and are often associated with complex history effects. Among these, APG TBLs are particularly complex yet important. 
As the pressure gradient increases and accumulates, the velocity profile becomes distorted, wall shear stress decreases, the boundary layer thickens, and may even separate. These changes can lead to energy loss, unsteady vortex shedding, and flow-induced vibrations. 
Therefore, accurately understanding the flow structure of APG TBLs, as well as predicting wall shear stress and separation locations, remains a significant challenge in turbulence modeling.

In the modeling of APG TBLs, accurately capturing the effects of PG and flow history presents a key challenge. APG TBLs evolve gradually in the streamwise direction.
As the flow develops, upstream influences accumulate and strongly affect downstream behavior. This poses a major difficulty in constructing a universal model that is applicable to various flow histories and pressure distributions.
In particular, for many years, researchers have preferred unified scaling laws based on local flow parameters, such as Reynolds numbers and local pressure-gradient coefficients. However, it is readily apparent that capturing the characteristics of an entire two-dimensional flow field using only a limited set of local parameters poses a formidable challenge.

This study focuses on this problem and aims to partially capture the history effect through local parameters. A new set of non-dimensional scalings is proposed, forming the foundation for constructing an estimation-correction model for the mean velocity profiles and total shear stress profiles in APG TBLs. The rest of the paper is organized as follows: 
The rest of \S\ref{sec:Introduction} introduces the concept of equilibrium flows and reviews recent research progress, with a particular focus on the model of wall-bounded turbulence proposed by \cite{Subrahmanyam_Cantwell_Alonso_2022}, which serves as the foundation of the present study. 
\S\ref{sec:Methodology} presents a new scaling method based on friction velocity and Clauser pressure gradient parameter, and then proposes an estimation–correction method founded on this scaling.
\S\ref{sec:Result and Discussion} demonstrates the validation results, and \S\ref{sec:Conclusion} discusses and summarizes the main findings.

\subsection{Equilibrium Flow}

Among all wall-bounded turbulent flows, some special cases are referred to as equilibrium flows or canonical flows. 
They exhibit streamwise self-similarity, meaning that all flow profiles can be collapsed onto a universal profile through appropriate scaling.
As a result, the governing partial differential equations can be reduced to ordinary differential equations, rendering the flow analytically tractable.
Clearly, in the outer region of an equilibrium TBL, the streamwise pressure distribution is not arbitrary, but must maintain a fixed balance with the wall shear stress.
The concept of equilibrium flow was first introduced by \cite{Clauser_1954}, who also demonstrated that a necessary and sufficient condition for a TBL to achieve equilibrium is that a non-dimensional PG parameter remains constant during its evolution from upstream to downstream. 
This parameter is known as Clauser pressure gradient parameter $\beta = (\delta_1/\tau_w)(\mathrm dp_e/\mathrm dx)$, where $\delta_1$ is the displacement thickness, $\tau_w$ is the wall shear stress, and $\mathrm dp_e/\mathrm dx$ is the streamwise gradient of the free-stream pressure. 
Since then, considerable attention has been devoted to self-similarity, and extensive research has been conducted on the scaling laws in equilibrium wall-bounded turbulence, as concluded comprehensively by \cite{Marusic_McKeon_Monkewitz_Nagib_Smits_Sreenivasan_2010} and \cite{William_Todd_2022}. 

Following Clauser’s formulation of the equilibrium criterion, \cite{Zagarola_Smits_1998} proposed a new scaling for TBLs, known as Zagarola-Smits scaling, which achieves better self-similarity in velocity profiles. A corresponding PG parameter was later introduced by \cite{Maciel_Rossignol_Lemay_2006} as $\beta _{ZS}$, used to characterize the local effect of PG.
Around the same time, \cite{Castillo_George_2001} also introduced a new PG parameter $\Lambda$ in TBLs to characterize the equilibrium states. They argue that $\Lambda$ takes three characteristic values, leading to three velocity deficit profiles.
Most recently, \cite{Wei_Knopp_2023} also proposed a new outer scaling based on maximum Reynolds shear stress location. 
The above new works have significantly advanced the characterization of detailed flow features, and have been increasingly adopted in recent studies. However, due to their complexity, the present theoretical analysis adopts the relatively simple Clauser pressure gradient parameter $\beta$ as the primary descriptor.

Two widely studied equilibrium cases of wall-bounded turbulence are turbulent channel flows and zero-pressure-gradient (ZPG) TBLs. 
Specifically, channel flows possess streamwise translational symmetry, and ZPG TBLs have a constant $\beta=0$, which clearly meets Clauser's equilibrium condition. 
Extensive studies on these two equilibrium flows provide a solid foundation for the present investigation of the more general, non-equilibrium, APG TBLs. 

For APG TBLs, the breakdown of symmetry and self-similarity renders the boundary layer governing equations non-analytical, and the flow exhibits a range of unique and complex features.
Among them, a key issue is that, the flow profiles can no longer be described solely by a set of local parameters, but instead must account for the entire history of the flow evolution, or the so-called history effects.
One of the manifestations is the delayed response to PG changes.
As \cite{William_Todd_2022} suggested, it takes several boundary layer thicknesses for an equilibrium boundary layer to adjust to PG changes. This conclusion is also consistent with the recovery distance observed by \cite{Volino_2020}.
In fact, \cite{Gungor_Gungor_Maciel_2024} offered a more intuitive illustration of history effect. They plotted profiles with the same non-dimensional PG ($\beta_{ZS}=0.1$), but both their velocity and Reynolds stress profiles differed significantly, clearly demonstrating the pronounced history effect in APG TBLs.

In recent years, considerable research has focused on the quantification of history effects.
For example, \cite{Vinuesa_Orlu_Sanmiguel_Ianiro_Discetti_Schlatter_2017} established a criterion to identify well-behaved APG TBLs, through which they successfully distinguished PG effects from history effects. 
Most recently, \cite{Gomez_McKeon_2025} utilized linear analysis to characterize PG history. 
Although these works have provided deep insight into quantifying history effects, a systematic predictive method remains lacking.

\subsection{Basic Method}

Based on the near-wall similarity, \cite{Subrahmanyam_Cantwell_Alonso_2022} developed a well-behaved universal method with minimal assumptions, which can accurately predict mean velocity profiles
and stress profiles for channel flows and ZPG TBLs, the two aforementioned equilibrium flows. 
In this section, we will briefly describe their pioneering work, which also forms the foundation for our subsequent research on non-equilibrium APG TBLs.

The study focuses on incompressible wall-bounded turbulent flows. The governing equations are the two-dimensional, statistically time-independent Reynolds-averaged Navier–Stokes (RANS) equation with boundary layer approximation:
\begin{equation}
    \frac{\partial u u}{\partial x}+ \frac{\partial u v}{\partial y}
    +\frac{\partial \langle u' v'\rangle}{\partial y}+\frac 1 \rho \frac{\mathrm d p_e}{\mathrm d x}-\nu \frac{\partial^2 u}{\partial y^2}=0,
    \label{eq:RANS}
\end{equation}
together with the continuity equation:
\begin{equation}
    \frac{\partial u}{\partial x}+\frac{\partial v}{\partial y}=0,
    \label{eq:Mass Conservation}
\end{equation}
where $x$ is the flow direction, $y$ the wall-normal direction, $(u,v)$ the streamwise and wall-normal velocity, $p_e$ the free-stream pressure, and $\rho,\nu$ the density and kinematic viscosity of the fluid, respectively. 
$\langle u' v'\rangle$ is the Reynolds shear stress, representing the transport of $x-$momentum along the $y$ direction by turbulence fluctuations. At the wall, the no-slip boundary condition goes:
\begin{equation}
    y=0:\quad u = v = 0.
    \label{eq:RANS_boundary_conditon}
\end{equation}

When the Reynolds number ($Re$) is sufficiently high and turbulence is fully developed, all flow structures in the near-wall regions, such as velocity streaks and quasi-streamwise vortices, are primarily generated by the wall boundary. 
Therefore, the velocity and stress profiles in the near-wall regions exhibit certain similarities under different flow conditions. Based on this, the local wall shear stress $\tau_w$ is adopted as the scaling basis, and the corresponding length scale and velocity scale are referred to as friction length $\delta_\nu$ and friction velocity $u_\tau$, respectively, which are defined as follows:
\begin{equation}
    \tau_w=\mu \left.\frac{\partial u}{\partial y}\right|_{y=0},\quad u_\tau =\sqrt{\frac{\tau_w }{\rho}},\quad \delta_\nu=\frac{\nu}{u_\tau}.
    \label{eq:friction_velocity_based_scaling}
\end{equation}
Using this unified set of wall scales, the non-dimensional scaling of the relevant flow quantities is as follows:
\begin{equation}
    y^+=\frac{y}{\delta_\nu},\quad
    (u^+,v^+) = \frac{(u,v)}{u_\tau},\quad \tau_{R}^+=\frac{-\langle u'v'\rangle}{u_\tau^2},\quad \tau^+=\tau^+_R+\frac{\partial u^+}{\partial y^+}  .
    \label{eq:friction_velocity_based_nondimensionalization}
\end{equation}
Here, $\tau_R^+$ denotes the non-dimensional Reynolds shear stress, and $\tau^+$ denotes the non-dimensional total shear stress, which includes both Reynolds stress and viscous stress. In the rest of the paper, quantities with a superscript "$+$" indicate values non-dimensionalized using the local wall friction scales.

With the basic equations, \cite{Subrahmanyam_Cantwell_Alonso_2022} start from one of the simplest equilibrium cases, the turbulent channel flow.
In this situation, since the mean flow field is independent of $x$, the convective term ($\partial u u/\partial x + \partial u v /\partial y$) vanishes:
\begin{equation}
    \frac{\mathrm d \langle u' v' \rangle}{\mathrm d y} +\frac 1 \rho \frac{\mathrm d p_e}{\mathrm d x}-\nu \frac{\mathrm d^2 u}{\mathrm d y^2}=0
    \label{eq:channel_flow},
\end{equation}
and the boundary condition at the outer edge goes:
\begin{equation}
    y=\delta:\quad \frac{\mathrm du}{\mathrm dy} = 0,
    \label{eq:channel_flow_boundary_condition}
\end{equation}
where $\delta$ is the half height of the channel. Combining \eqref{eq:channel_flow}, \eqref{eq:channel_flow_boundary_condition}, and \eqref{eq:friction_velocity_based_nondimensionalization}, and integrating the equation with respect to $y^+$, we have:
\begin{equation}
   \tau_{R}^++\frac{\mathrm d u^+}{\mathrm d y^+}=\tau^+ =1 - \frac{y^+}{Re_\tau},
   \label{eq:channel_flow_integral_once}
\end{equation}
where $Re_\tau = u_\tau \delta /\nu$ is the friction Reynolds number. In order to close the equation,  the shear stress is modeled using classical mixing length theory \citep{Von_1931,Van_1956}:
\begin{equation}
   \tau_{R}^+=\left(\lambda^+(y^+)\frac{\partial u^+}{\partial y^+} \right)^2,
   \label{eq:mixing_length_theory}
\end{equation}
where $\lambda^+(y^+)$ is the mixing length. Combining \eqref{eq:channel_flow_integral_once} and \eqref{eq:mixing_length_theory}, the result is a quadratic equation with respect to $\mathrm d u^+/\mathrm d y^+$. The positive root is as follows:
\begin{equation}
   \frac{\mathrm d u^+}{\mathrm d y^+} = \frac{-1+\sqrt{1+4(\lambda^+(y^+))^2(1-y^+/Re_\tau)}}{2(\lambda^+(y^+))^2}.
   \label{eq:channel_flow_du/dy}
\end{equation}
Simply integrate it from the wall to any reference location, the $u^+$ profile is acquired, as a function of $y^+$ and $Re_\tau$:
\begin{equation}
   u^+(y^+, Re_\tau)= \int_0^{y^+} \frac{-1+\sqrt{1+4(\lambda^+(s))^2(1-s/Re_\tau)}}{2(\lambda^+(s))^2}\mathrm ds.
   \label{eq:channel_flow_u(y)}
\end{equation}

The mixing length model used here is introduced by \cite{Cantwell_2019} to approximate pipe flows:
\begin{equation}
   \lambda^+(y^+)=ky^+\frac{1-e^{-(y^+/a)^m}}{\left(1+\left(\frac{y^+}{bRe_\tau}\right)^n\right)^{1/n}}.
   \label{eq:lambda}
\end{equation}
This model consists of five empirically determined parameters $k,a,m,b,n$. 
The constant $k$ is closely related to the Kármán constant $\kappa$ in the logarithmic law. 
The parameter $a$ represents a wall-damping length scale which is first proposed by \cite{Van_1956}, and parameter $m$ determines the rate of damping. 
Together, $k, a, m$ define the overall extent of the wall layer, spanning from the viscous sublayer through the buffer layer to the end of the logarithmic region. 
The outer flow is accounted for in the denominator of \eqref{eq:lambda} which includes a length scale $b$ that indicates the layer thickness, as well as an exponent $n$ that helps shape the outer profile.

Technically, the parameters $(k,a,m,b,n)$ of the mixing length model \eqref{eq:lambda} are determined by minimizing the mean squared error (MSE) using a cost function:
\begin{equation}
   L(Re_\tau; k,a,m,b,n)=\frac 1N \sum_{i=1}^N (u_{(i)}^+-u^+(y_{(i)}^+,Re_{\tau };k,a,m,b,n))^2,
   \label{eq:channel_flow_cost_function}
\end{equation}
where the summation is over all training set data points, and $N$ is the total number of points. Basically, in the original method, \cite{Subrahmanyam_Cantwell_Alonso_2022} performed the MSE optimization for every profile separately. Therefore, in equation \eqref{eq:channel_flow_cost_function}, all data points share the same $Re_\tau$, but correspond to different wall-normal locations. The subscript $(i)$ indicates the $i^{\mathrm {th}}$ point: $u_{(i)}^+$ is the experiment or DNS result at $y_{(i)}^+$, and $u^+(y_{(i)}^+,Re_{\tau};k,a,m,b,n)$ is the prediction of the model \eqref{eq:channel_flow_u(y)} at the given $y_{(i)}^+$. Through minimizing the cost function $L(Re_\tau; k,a,m,b,n)$, the optimal parameters $(k,a,m,b,n)$ can be acquired at a given $Re_\tau$.

At this point, the modeling of channel flows is complete. In the following, the same method will be applied to the boundary layer. However, in this case, the convection term doesn't vanish and the governing equation\eqref{eq:RANS} remains non-integrable. 
To address this issue, the following factors are considered: First, since the flow in the inner region is primarily controlled by the wall, all wall-bounded turbulent flows exhibit high similarity in their near-wall velocity profiles when scaled as equation \eqref{eq:friction_velocity_based_nondimensionalization}. Second, turbulent channel flows and TBLs share the same boundary condition at the outer edge, which is $\partial u/\partial y=0$.
Therefore, as \cite{Subrahmanyam_Cantwell_Alonso_2022} did, the velocity profile \eqref{eq:channel_flow_u(y)} for channel flows is directly adopted in general boundary layer cases, with empirical parameters $k,a,m,b,n$ specially modified. The strategy for parameter adjustment here also relies on minimizing the mean squared error, which is entirely consistent with equation \eqref{eq:channel_flow_cost_function}.

The only difference is that $\delta$ stands for the nominal thickness of the boundary layer, which is the distance from the wall to the point at which $u$ reaches 99\% of free-stream velocity $u_e$. 
For ZPG TBLs over a wide range of Reynolds numbers, this method not only predicts the mean velocity profiles accurately, but also allows for the derivation of the Reynolds shear stress distribution by substituting the velocity expression into the Kármán momentum integral \citep{Karman_1921,Gruschwitz_1931}, and converting the streamwise derivative with respect to $x$ into a derivative with respect to the friction Reynolds number $Re_\tau$. The result is as follows:
\begin{equation}
   \tau_{R}^+= 1 -\frac{\partial u^+}{\partial y^+}-\frac{u^+ \frac{\partial}{\partial Re_\tau}\int_0^{y^+}u^+ \mathrm d y^+ -\frac{\partial}{\partial Re_\tau}\int_0^{y^+}u^{+2}\mathrm d y^+-u_e^+\frac{\mathrm d u_e^{+-1}}{\mathrm dRe_\tau}\int_0^ {y^+}  u^{+2}\mathrm d y^+}{u_e^+ \frac{\mathrm d}{\mathrm d Re_\tau}\int_0^{Re_\tau}u^+ \mathrm d y^+ -\frac{\mathrm d}{\mathrm dRe_\tau}\int_0^{Re_\tau}u^{+2}\mathrm d y^+-u_e^+\frac{\mathrm d u_e^{+-1}}{\mathrm dRe_\tau}\int_0^ {Re_\tau}  u^{+2}\mathrm d y^+},
   \label{eq:ZPGTBL_tau}
\end{equation}
where $u_e^+ = u_e/u_\tau$, which can also be expressed explicitly by \eqref{eq:channel_flow_u(y)}:
\begin{equation}
   u_e^+(Re_\tau)= \int_0^{Re_\tau} \frac{-1+\sqrt{1+4(\lambda^+(s))^2(1-s/Re_\tau)}}{2(\lambda^+(s))^2}\mathrm ds.
   \label{eq:channel_flow_ue}
\end{equation}

The stress profile \eqref{eq:ZPGTBL_tau} is also accurate and useful for ZPG TBLs. However, when trying to apply this method to non-equilibrium APG TBL cases, the model \eqref{eq:channel_flow_u(y)} and \eqref{eq:ZPGTBL_tau} don't fit well even with its best manually adjusted parameters $k,a,m,b,n$. 
Additionally, because of the loss of streamwise self-similarity, even in the same flow field, flow structures at different streamwise locations have typically little  dependency on each other.
Therefore, the velocity profile, stress profile, or the mixing length profile, are all generally irrelevant at different streamwise locations, which means the parameters have to be re-determined each time the formula \eqref{eq:channel_flow_u(y)} or \eqref{eq:ZPGTBL_tau} is used, as seen in Table 5 of \cite{Subrahmanyam_Cantwell_Alonso_2022}.
As a result, to some extent, this method has lost its effect.

\section{Methodology}\label{sec:Methodology}

\subsection{New Scaling}\label{subsec: Non-dimensionalization}

As discussed above, one of the key difficulties of APG TBLs lies in the loss of self-similarity along $x$ direction due to history effects. 
The streamwise development of the flow in an integral form can be shown in the Kármán momentum integral \citep{Karman_1921,Gruschwitz_1931}:
\begin{equation}
    \frac{\mathrm d}{\mathrm dx}(u_e^2 \theta) = u_\tau^2 (1+\beta),
   \label{eq:Karman_momentum_integral}
\end{equation}
where $\theta$ is the momentum thickness, and $\beta$ is Clauser pressure gradient parameter.
The left-hand side of \eqref{eq:Karman_momentum_integral} is the total derivative with respect to $x$, which clearly indicates the streamwise development of the total shape of the boundary layer. 
Therefore, as an substitute for classic $u_\tau$, we propose to use
\begin{equation}
   u_{\beta} = u_\tau \sqrt{1+\beta}
   \label{eq:u_star}
\end{equation}
as the basic scale for velocity. This new scaling combines wall shear stress, viscous stress and PG effect, and provides a potential approach to restoring the self-similarity. Note that, since our discussion is limited to APG TBL cases, corresponding to $\beta>0$, the square root operation is valid.
The corresponding length scale and Reynolds number are as follows:
\begin{equation}
    \delta_\beta=\frac{\nu}{u_\beta},\quad
    Re_\tau^* = \frac{u_\beta \delta}{\nu}.
    \label{eq:y_star}
\end{equation}
Under this scaling, the new non-dimensional quantities are denoted by the superscript $*$, and their specific forms are as follows:
\begin{equation}
    y^*=\frac{y}{\delta_\beta},\quad
    (u^*,v^*) = \frac{(u,v)}{u_\beta},\quad \tau_{R}^*=\frac{-\langle u'v'\rangle}{u_\beta^2},\quad \tau^*=\tau^*_R+\frac{\partial u^*}{\partial y^*}.
    \label{eq:u_star_based_nondimensionalization}
\end{equation}

\subsection{Estimation-Correction Method}\label{subsec:Iteration method}

Another mismatch introduced by Subrahmanyam \etal 's (\citeyear{Subrahmanyam_Cantwell_Alonso_2022})
original method is the formula itself. As previously noted, the original model directly applied the velocity profile of channel flows to TBLs, which inevitably leads to discrepancies.
Based on the new scaling method \eqref{eq:u_star_based_nondimensionalization}, the accurate form of velocity profile in boundary layer cases is:
\begin{equation}
   u^*\left(y^*, Re_\tau^*,\beta,\frac{\mathrm {d}\beta}{\mathrm {d} x},\frac{\mathrm {d}^2\beta}{\mathrm {d} x^2},...\right)= \int_0^{y^*} \frac{-1+\sqrt{1+4(\lambda^*(s))^2\tau^*\left(s,Re_\tau^*,\beta,\frac{\mathrm {d}\beta}{\mathrm {d} x},\frac{\mathrm {d}^2\beta}{\mathrm {d} x^2},...\right)}}{2(\lambda^*(s))^2}\mathrm ds.
   \label{eq:u(y)}
\end{equation}
Here, the modeling of $\lambda^*$ is the same as equation \eqref{eq:lambda}; it only requires replacing all superscripts $^+$ with $^*$:
\begin{equation}
   \lambda^*(y^*,Re_\tau)=ky^*\frac{1-e^{-(y^*/a)^m}}{\left(1+\left(\frac{y^*}{bRe_\tau}\right)^n\right)^{1/n}}.
   \label{eq:lambdastar}
\end{equation}

Comparing \eqref{eq:channel_flow_u(y)} and \eqref{eq:u(y)}, in addition to the modification of the scaling, the term $1-s/Re_\tau$ in equation \eqref{eq:channel_flow_u(y)}, which is the total shear stress in channel flows, has been replaced by the more general expression $\tau^*$. 
In boundary layer cases, the exact form of $\tau^*$ is certainly not available. As an improvement to the original approach, we still fully account for the near-wall similarity, but instead of directly applying the channel flows results, we consider an estimation-correction method, using the linear shear stress distribution \eqref{eq:channel_flow_integral_once} from channel flow as the initial estimation:
\begin{equation}
    \widetilde \tau^{*}=\frac{1}{1+\beta}\left(1- \frac{y^*}{Re_\tau^*}\right).
   \label{eq:tau0}
\end{equation}
Here, and throughout the whole paper, variables with upward tildes represent estimated quantities.
Next, we will construct a correction for $\tau^*$ by solving the original RANS equations, using the estimated $\widetilde u^*$ as an intermediate variable. According to equation \eqref{eq:u(y)}, the estimated $\widetilde u^*$ can be readily expressed as 
\begin{equation}
    \widetilde u^*(y^*, Re_\tau^*,\beta)= \int_0^{y^*} \frac{-1+\sqrt{1+4(\lambda^*(s))^2\frac{1}{1+\beta}\left(1- \frac{s}{Re_\tau^*}\right)}}{2(\lambda^*(s))^2}\mathrm ds.
   \label{eq:u1}
\end{equation}

By substituting the given $\widetilde u^*$ into the RANS equations, and then sequentially expanding all the terms and eliminating $v^*$ by the continuity equation, we can eventually solve for the modified
$\tau^*$ and $u^*$.
The most challenging part lies in handling the convective term $u\partial u/\partial x$, since our model does not provide an explicit expression of the estimated velocity $\widetilde u^*$ as a function of $x$. To address this issue, we naturally assume that $\widetilde u^*$ as a function of $(y^*, Re_\tau^*,\beta)$, is universal regardless of streamwise location, which means $\widetilde u^*$ isn't explicitly dependent on $x$. Therefore, $\partial \widetilde u^*/\partial x$ can be converted into the derivatives of other flow parameters with respect to $x$:
\begin{equation}
    \frac{\partial}{\partial x}\widetilde u^*(y^*, Re_\tau^*,\beta) 
    = \frac{\partial \widetilde u^*}{\partial y^*}\frac{\partial y^*}{\partial x}
    + \frac{\partial \widetilde u^*}{\partial Re_\tau^*}\frac{\mathrm{d} Re_\tau^*}{\mathrm{d} x}
    + \frac{\partial \widetilde u^*}{\partial \beta^*}\frac{\mathrm{d} \beta^*}{\mathrm{d} x}.
   \label{eq:expansion example}
\end{equation}

According to equation \eqref{eq:u1} and \eqref{eq:lambdastar}, $\partial \widetilde u^*/\partial y^*$, $\partial \widetilde u^*/\partial Re_\tau^*$ and $\partial \widetilde u^*/\partial \beta$ are easy to acquire. 
Meanwhile, given the definition equations 
\eqref{eq:friction_velocity_based_scaling}, \eqref{eq:u_star}, \eqref{eq:y_star} and \eqref{eq:u_star_based_nondimensionalization}, $\partial y^*/\partial x$ can be expanded in terms of $\mathrm dRe_\tau^*/\mathrm dx$ and $\mathrm d\beta/\mathrm d x$. 
Since $\beta$ is generally treated as a known quantity, its local derivative $\mathrm d\beta/\mathrm d x$ can also be readily obtained from the outer edge boundary conditions. However, $\mathrm dRe_\tau^*/\mathrm dx$ cannot be assumed known, as it depends on the evolving inner-layer structures.

In order to solve this problem, and in order to introduce more integral information, we go back to the expanded form of Kármán momentum integral \eqref{eq:Karman_momentum_integral}:
\begin{equation}
    \frac{1}{u_e^{*2}}\left(1+\frac{2\beta}{1+\beta}\frac{\theta}{\delta_1}\right)=\frac{\mathrm d\theta}{\mathrm dx}=\frac{\partial \theta}{\partial \beta}\frac{\mathrm d\beta}{\mathrm dx}+\frac{\partial\theta }{\partial Re_\tau^*}\frac{\mathrm d Re_\tau^*}{\mathrm d x}+\frac{\partial \theta}{\partial u_e}\frac{\mathrm d u_e}{\mathrm d x}+\frac{\partial \theta}{\partial u_e^*}\left( \frac{\partial u_e^*}{\partial Re_\tau^*}\frac{\mathrm dRe_\tau^*}{\mathrm dx}+\frac{\partial u_e^*}{\partial \beta}\frac{\mathrm d \beta}{\mathrm dx}\right),
   \label{eq:Karman_momentum_integral_expanded}
\end{equation}
where the momentum thickness $\theta=({\nu}/{u_e})\int_0^{Re_\tau^*}\widetilde u^*(1-\widetilde u^*/u_e^*)\mathrm ds$ is a function of $(Re_\tau^*, \beta,u_e^*,u_e)$. 
Note that $u_e$ is actually connected with the pressure gradient by the equilibrium equation outside the boundary layer:
\begin{equation}
    u_e
    \frac{\mathrm d u_e}{\mathrm dx}=-\frac 1 \rho \frac{\mathrm dp}{\mathrm dx}=\frac{-1}{\rho \nu}\frac{\beta \tau_w u_\tau^*}{\delta_1^*}.
    \label{eq:ue_with_beta}
\end{equation}
Combine equation \eqref{eq:Karman_momentum_integral_expanded} and \eqref{eq:ue_with_beta}, and eliminate $\mathrm du_e/\mathrm dx$, we can get the derivative relationship between $Re_\tau^*$ and $x$:
\begin{equation}
    \frac{\mathrm d Re_\tau^*}{\mathrm dx}=\frac{u_\beta}{\nu}\frac
    {\frac{1}{u_e^{*}}\left(1+\frac{\beta}{1+\beta}\frac{\theta^*}{\delta_1^*}\right)-\gamma \frac{\partial u_e^*}{\partial \beta}\int_0^{Re_\tau^*}\frac{\widetilde u^{*2}}{u_e^{*2}}\mathrm dy^*-\gamma \int_0^{Re_\tau^*}\frac{\partial \widetilde u^*}{\partial \beta}\left(1-\frac{2\widetilde u^*}{u_e^*}\right)\mathrm d y^*}{\int_0^{Re_\tau^*}\frac{\partial \widetilde u^*}{\partial Re_\tau^*}\mathrm dy^* -\int_0^{Re_\tau^*}\frac{\partial \widetilde u^{*2}}{u_e^*\partial Re_\tau^*}\mathrm dy^*+\frac{\partial u_e^*}{u_e^{*2}\partial Re_\tau^*}\int_0^{Re_\tau^*}\widetilde u^{*2}\mathrm dy^*},
    \label{eq:dRetau_dx}
\end{equation}
where $\delta_1^*$ and $\theta^*$ are the non-dimensional form of $\delta_1$ and $\theta$, based on the length scale $\delta_\beta$, and $u_e^*$ is the estimated non-dimensional freestream velocity, which can be expressed explicitly by equation \eqref{eq:u1}:
\begin{equation}
    u_e ^* = \widetilde u^*(y^*=Re_\tau^*, Re_\tau^*,\beta)= \int_0^{Re_\tau^*} \frac{-1+\sqrt{1+4(\lambda^*(s))^2\frac{1}{1+\beta}\left(1- \frac{s}{Re_\tau^*}\right)}}{2(\lambda^*(s))^2}\mathrm ds.
   \label{eq:uestar}
\end{equation}

In equation \eqref{eq:dRetau_dx}, a newly-defined non-dimensional parameter is 
\begin{equation}
    \gamma =\frac \nu{u_\beta}\frac{\mathrm d\beta}{\mathrm dx},
    \label{eq:Gamma Definition}
\end{equation}
representing the second order derivative of pressure in the streamwise direction. 
However, a higher order spatial derivative term will never be obtained from the RANS equations through time-averaging or integration. This means $\gamma$ captures additional information beyond RANS equations, and therefore, we consider it to partially represent the higher-order local history effect of PG. 
Interestingly, the effect of PG derivative on velocity profile in the form of $\mathrm d\beta/\mathrm dx$ is also plotted by \cite{Gungor_Gungor_Maciel_2024}. 

Further discussion of $\gamma$ will be conducted in Section \ref{sec:Result and Discussion}, but at this point, we're able to handle the term $\partial u/\partial x$ in equation \eqref{eq:Mass Conservation}:
\begin{equation}
    \begin{aligned}
    \frac{\partial v^*}{\partial y^*}=&-\frac{\nu}{u_\beta^2}\frac{\partial u}{\partial x}
    \\
    =
    &-\gamma \frac{\partial \widetilde u^*}{\partial \beta}+\frac{\beta}{(1+\beta)u_e^{*2}\delta_1^*}\left(\widetilde u^*+\frac{\partial \widetilde u^*}{\partial y^*}y^*\right)+\frac{1}{u_e^*}\gamma\frac{\partial u_e^*}{\partial \beta}\left(\widetilde u^*+\frac{\partial \widetilde u^*}{\partial y^*}y^*\right)
    \\
    &-\frac{\nu}{u_\beta}\frac{\mathrm dRe_\tau^*}{\mathrm dx}\left(
    \frac{\partial \widetilde u^*}{\partial Re_\tau^*}-\frac 1 {u_e^*}\frac{\partial u_e^*}{\partial Re_\tau^*}\left(\widetilde u^*+\frac{\partial \widetilde u^*}{\partial y^*}y^*\right)
    \right).
    \end{aligned}
    \label{eq:v(y)}
\end{equation}
Simply integrate equation \eqref{eq:v(y)}, and an explicit expression of $v^*$ as a function of $y^*$ is figured out. 
Using the same method, the convective term $\partial uu/\partial x$ in equation \eqref{eq:RANS} can be handled. 
After substituting $v^*$, the final expression for the corrected $\tau^*$ can be obtained:
\begin{gather}
    \begin{aligned}
    \tau^*=&\frac{1}{\beta+1}+\frac{\beta y^*}{(1+\beta)\delta_1^*}-\left( \frac{\beta}{(1+\beta)\delta_1^*}+u_e^*\gamma\frac{\partial u_e^*}{\partial \beta} \right)\int_0^{y^*}\frac{\widetilde u^{*2}}{u_e^{*2}}\mathrm dy^*
    \\
    &-G(y^*)\left(
    1+\frac{\beta}{1+\beta}\frac{\theta^*}{\delta_1^*}-\gamma u_e^*\left(\frac{\partial u_e^*}{\partial \beta}\int_0^{Re_\tau^*}\frac{\widetilde u^{*2}}{u_e^{*2}}\mathrm dy^*+\int_0^{Re_\tau^*}\frac{\partial \widetilde u^*}{\partial \beta}\left(1-\frac{2 \widetilde u^*}{u_e^*}
    \right)\mathrm d y^*
    \right)
    \right)
    \\
    &+\gamma\left(
    2\int_0^{y^*}\widetilde u^*\frac{\partial \widetilde u^*}{\partial \beta}\mathrm d y^* -\widetilde u^*\int_0^{y^*}\frac{\partial \widetilde u^*}{\partial \beta}\mathrm d y^*
    \right),
    \end{aligned}
    \label{eq:tau1_y}
    \\
    \mathrm{where} \quad G(y^*)= \frac{\widetilde u^* \frac{\partial}{\partial Re_\tau^*}\int_0^{y^*}\widetilde u^* \mathrm d y^* -\frac{\partial}{\partial Re_\tau^*}\int_0^{y^*}\widetilde u^{*2}\mathrm d y^* -u_e^*\frac{\partial u_e^{*-1}}{\partial Re_\tau^*}\int_0^ {y^*}  \widetilde u^{*2}\mathrm d y^*}{u_e^* \frac{\partial}{\partial Re_\tau^*}\int_0^{Re_\tau^*}\widetilde u^* \mathrm d y^* -\frac{\partial}{\partial Re_\tau^*}\int_0^{Re_\tau^*}\widetilde u^{*2}\mathrm d y^* -u_e^*\frac{\partial u_e^{*-1}}{\partial Re_\tau^*}\int_0^ {Re_\tau^*}\widetilde u^{*2}\mathrm d y^*}.
    \nonumber
\end{gather}

At this point, we have obtained an explicit expression for $\tau^*$.
Comparing equation (\ref{eq:tau1_y}) with (\ref{eq:ZPGTBL_tau}) which is the result for ZPG TBLs in the original method \citep{Subrahmanyam_Cantwell_Alonso_2022}, it is straightforward to observe that when $\beta$ is identically zero, $\tau^*$ reduces to $\tau^+$ in ZPG TBLs. 

Immediately, based on the accurate relation between $u^*$ and $\tau^*$ in equation \eqref{eq:u(y)} the modified $u^*$ can then be determined as:
\begin{equation}
   u^*\left(y^*, Re_\tau^*,\beta,\gamma\right)= \int_0^{y^*} \frac{-1+\sqrt{1+4(\lambda^*(s))^2\tau^*\left(s,Re_\tau^*,\beta,\gamma\right)}}{2(\lambda^*(s))^2}\mathrm ds.
   \label{eq:u2(y)}
\end{equation}

Just as in the original method, the parameters $(k,a,m,b,n)$ in $\lambda^*$ of equation \eqref{eq:lambdastar} need to be determined by minimizing the mean square error. 
Similar to equation \eqref{eq:channel_flow_cost_function}, the cost function here is defined as:
\begin{equation}
   L(k,a,m,b,n)=\frac 1{MN} \sum_{j=1}^M \sum_{i=1}^N (u_{(i,j)}^*-u^*(y_{(i,j)}^*,Re_{\tau (j)}^*,\beta _{(j)},\gamma_{(j)}; k,a,m,b,n)^2.
   \label{eq:APGTBL_cost_function}
\end{equation}
Here, the training set data points are not limited to one single profile, but are potentially obtained from different streamwise locations and even different flow fields. $M$ denotes the number of profiles, and $N$ denotes the number of data points in each profile. Therefore, there are $MN$ data points in total. 
The subscript $(j)$ indicates the $j^{\mathrm {th}}$ profile, and subscript $(i)$ indicates the $i^{\mathrm {th}}$ data point. $u_{(i,j)}^*$ is the experiment or DNS result at $y_{(i,j)}^*$, and $u^*(y_{(i,j)}^*,Re_{\tau (j)}^*,\beta _{(j)},\gamma_{(j)}; k,a,m,b,n)$ is the prediction of the corrected velocity model \eqref{eq:u2(y)} given $(y^*_{(i,j)},Re_{\tau, (j)},\beta _{(j)},\gamma_{(j)})$.
Since profiles with different $Re_\tau$, $\beta$ or $\gamma$ are jointly included in the MSE optimization, the resulting optimal parameters $(k,a,m,b,n)$ can be regarded as universal.

Using flow field b1n (See Table \ref{tab:flow_description}) as the training dataset, we obtained the optimal parameters shown in Table\ref{tab:Optimal_parameters}. It is shown that the optimal $k=0.451$, which to some extent represents the Kármán constant $\kappa$, is larger than $\kappa=0.384$ obtained in Princeton Superpipe (PSP) data by \cite{Nagib_Chauhan_2008}, $\kappa = 0.37$ given by high-Reynolds-number LES of \cite{Inoue_Pullin_2011}, and $\bar k=0.4169$ given by \cite{Subrahmanyam_Cantwell_Alonso_2022} in his tested APG TBL cases. 

\begin{table}
  \begin{center}
\def~{\hphantom{0}}
  \begin{tabular}{cccccc}
        parameter  & $k$ & $a$ & $m$ & $b$ & $n$ \\[3pt]
        Optimal value & $0.451$ & $22.7$ & $1.90$ & $0.166$ & $15.0$ \\ 
  \end{tabular}
  \caption{Optimal parameters obtained using flow field b1n as the training set.}
  \label{tab:Optimal_parameters}
  \end{center}
\end{table}

Another issue to discuss is that due to the influence of PG, the flow outside the boundary layer may not stably reach the free stream velocity $u_e$. 
Therefore, in order to determine the nominal thickness $\delta$, a diagnostic function is used here:
\begin{equation}
   f(y^*)=y^*\frac{\partial u^*}{\partial y^*},
   \label{eq:diagnostic_function}
\end{equation}
and the outer edge of the boundary layer is defined as the location where $f(y^*)$ drops to $1/n$ of its second peak value. 
Conventionally, $n$ is taken as $2$, so that this location corresponds to the definition of $\delta_{99}$ in ZPG TBLs approximately. 
However, in this case, it is desirable that the Reynolds shear stress $\tau_{R}^*$ to be close to zero at the outer edge of the boundary layer, requiring a slight expansion of the boundary layer thickness. Experiments have shown that a better definition is to use $\delta_{995}$ \citep{Subrahmanyam_Cantwell_Alonso_2022}. Accordingly, $n$ is set to $5$ in the diagnostic function criterion.

\section{Result}\label{sec:Result and Discussion}
\subsection{Database}

The model is validated against eight DNS or LES flow fields, with the overall ranges of parameters presented in table \ref{tab:flow_description}, where $Re_\theta = {u_e \theta}/{\nu} $ denotes the momentum thickness Reynolds number. 
Among them, flow m18n has the strongest PG, and therefore, the strongest history effect and local non-equilibrium effect, while flow b14 has the highest Reynolds number. 
The tested flow fields cover a wide range of cases, with Clauser pressure gradient $\beta$ reaching a maximum of $4.53$, $Re_\theta$ peaking at $9.65\times 10^3$, and the second-order pressure gradient $\gamma$ ranging from $-4.57\times 10^{-3}$ to $2.82\times 10^{-4}$. 
Flow data LSM2017 is incomplete, and therefore only the numerical results within its equilibrium region are used for validation.

\begin{figure}[htbp]
    \centering
    \includegraphics[width=\textwidth]{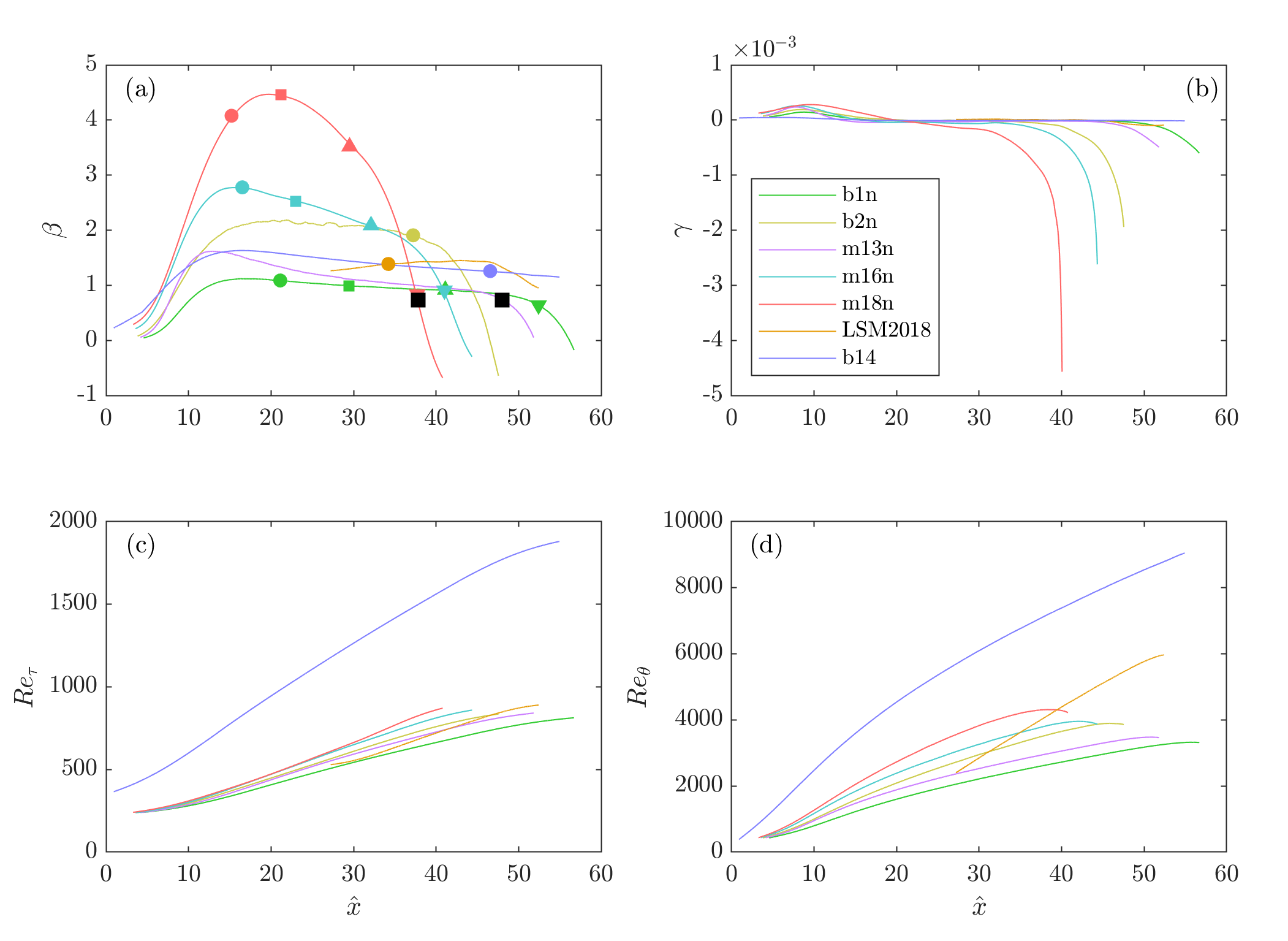}
    \caption{An overall description of seven flow fields along streamwise direction. The data are obtained directly through DNS or LES \citep{Bobke_Vinuesa_Örlü_Schlatter_2017,Yoon_Hwang_Sung_2018,Pozuelo_Li_Schlatter_Vinuesa_2022}. The horizontal axis denotes the streamwise coordinate $\hat x=x/\langle \delta_{99}\rangle$, where $x$ is non-dimensionalized by the nominal boundary layer thickness $\delta_{99}$ averaged over the computational domain. \\
    (a) Clauser pressure gradient $\beta$; (b) Streamwise derivative of pressure gradient $\gamma$ as defined in \eqref{eq:Gamma Definition};\\
    (c) Friction Reynolds number $Re_\tau$; (d) Momentum thickness Reynolds number $Re_\theta$;\\
    The markers in subfigure (a) indicate the locations of the profiles used for validation in Section \ref{subsec:Iteration Model}. 
    }
    \label{fig: Overall_profile}
\end{figure}

The streamwise development of flow parameters is illustrated in Figure \ref{fig: Overall_profile}. Here and throughout the rest of the paper, streamwise location $x$ is non-dimensionalized by the mean nominal boundary layer thickness $\delta_{99}$ within the computational domain, $\hat x=x/\langle \delta_{99}\rangle$. 
As shown in Figure \ref{fig: Overall_profile}(a), Clauser pressure gradient $\beta$ in all flow fields undergoes a process of first increasing and then decreasing. Correspondingly, in Figure \ref{fig: Overall_profile}(b), $\gamma$ is first positive and then negative. Negative $\gamma$ can reach very large negative values, with its absolute magnitude far exceeding the maximum positive values. This is due to the smaller wall friction, and therefore smaller $u_\tau$ in the downstream region, which also indicates that downstream regions are more susceptible to history effects. 
Figure \ref{fig: Overall_profile}(c) and (d) illustrate that the streamwise evolution of Reynolds number is relatively uniform and is not strongly related to variations of PG.

The markers in Figure \ref{fig: Overall_profile}(a) indicate the locations of the profiles used for validation in Section \ref{subsec:Iteration Model}, which span a broad parameter space. However, limited validation exists for regions with $\gamma>0$, due to the fact that the flow fields in these regions may have relatively low Reynolds numbers and incomplete turbulence development.

\begin{table}
  \begin{center}
\def~{\hphantom{0}}
  \begin{tabular}{ccccc}
       flow  & $\beta$ & $\gamma\times10^4$ &$Re_\theta\times10^{-2}$ & Reference \\[3pt]
        b1n & $0.05\sim1.12$ & $-6.00\sim 1.46$ & $4.27\sim33.1$ & \cite{Bobke_Vinuesa_Örlü_Schlatter_2017} \\ 
        b2n & $0.09\sim2.19$ & $-19.38\sim 1.91$ & $4.28\sim38.6$ & \cite{Bobke_Vinuesa_Örlü_Schlatter_2017} \\ 
        m13n & $0.06\sim1.62$ & $-4.91\sim 2.38$ & $4.34\sim34.8$ & \cite{Bobke_Vinuesa_Örlü_Schlatter_2017} \\ 
        m16n & $0.11\sim2.80$ & $-26.18\sim 2.61$ & $4.35\sim39.5$ & \cite{Bobke_Vinuesa_Örlü_Schlatter_2017} \\ 
        m18n & $0.14\sim4.53$ & $-45.65\sim 2.82$ & $4.34\sim43.1$ & \cite{Bobke_Vinuesa_Örlü_Schlatter_2017} \\ 
        LSM2017 & $0.23\sim2.57$ & $\sim0$ & $15.6$ & \cite{Lee_2017} \\ 
        LSM2018 & $0.85\sim1.55$ & $-0.98\sim 0.17$ & $24.0\sim59.6$ & \cite{Yoon_Hwang_Sung_2018} \\ 
        b14 & $0.02\sim1.65$ & $-0.40\sim 0.50$ & $2.36\sim96.5$ & \cite{Pozuelo_Li_Schlatter_Vinuesa_2022} \\
  \end{tabular}
  \caption{A brief description of eight flow fields, including Clauser pressure gradient $\beta$, streamwise derivative of pressure gradient $\gamma$ as defined in \eqref{eq:Gamma Definition}, and momentum thickness Reynolds number $Re_\theta$.}
  \label{tab:flow_description}
  \end{center}
\end{table}

\subsection{Verification of Similarity}

This section demonstrates that the new scaling method in Section \ref{subsec: Non-dimensionalization} effectively captures both self-similarity along the streamwise direction within a single flow field, and the overall similarity across different flow fields. 

\begin{figure}[htbp]
    \centering
    \includegraphics[width=\textwidth]{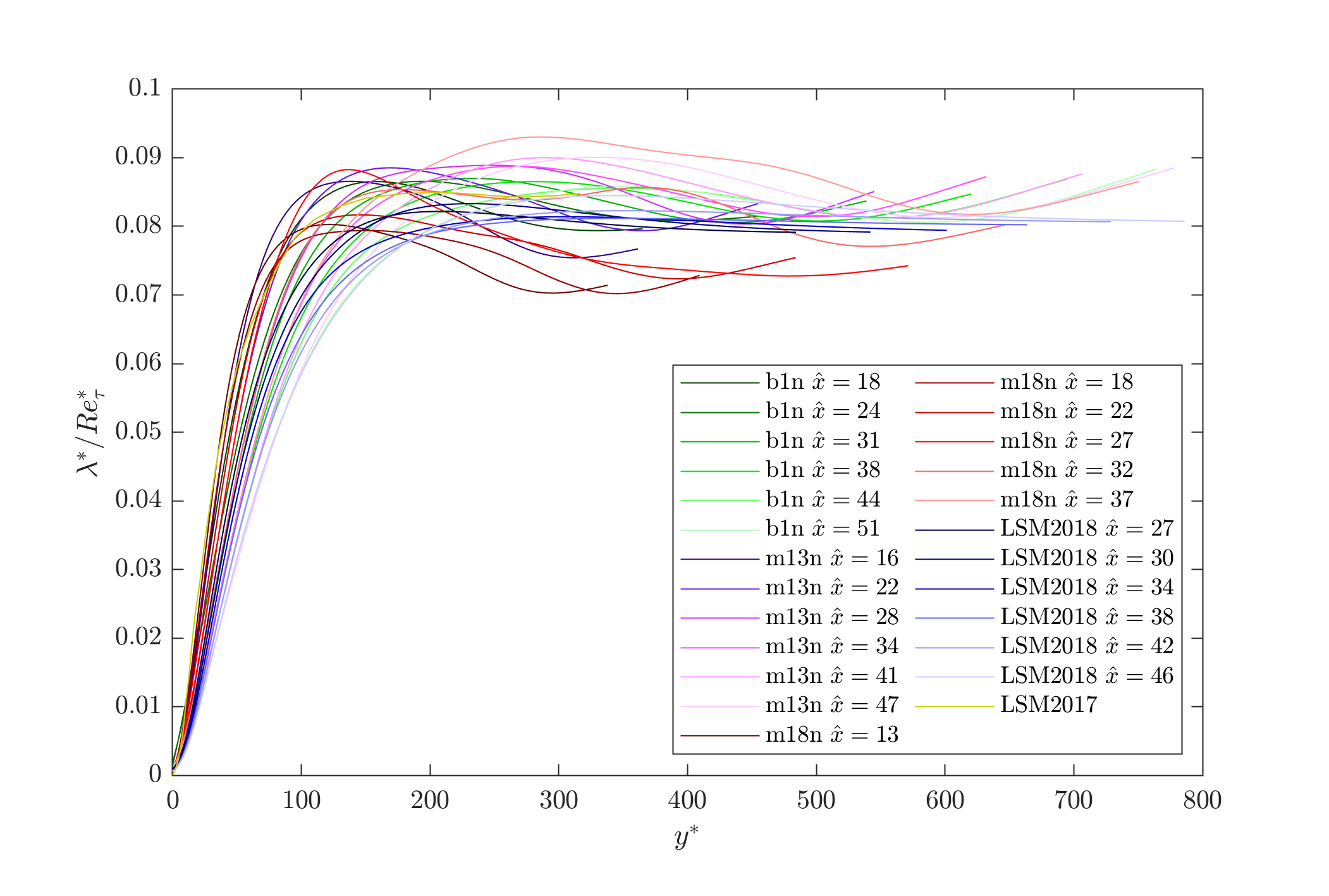}
    \caption{Profile of non-dimensional mixing length $\lambda^*$ divided by the friction Reynolds number $Re_\tau^*$ at different streamwise locations, in different flow fields. The data are obtained directly through DNS or LES \citep{Bobke_Vinuesa_Örlü_Schlatter_2017,Lee_2017,Yoon_Hwang_Sung_2018}. The result has been processed with Gaussian smoothing, and the critical ratio $n$ in criteria \eqref{eq:diagnostic_function} is set to $2$. Within the same flow field, lines of lighter color indicate locations further downstream.}
    \label{fig: Mixing_Length}
\end{figure}

As the non-dimensional mixing length model \eqref{eq:lambdastar} plays a fundamental role in the overall method, we calculate $\lambda^*$ directly from flow fields according to \eqref{eq:mixing_length_theory} (replacing the superscripts $^+$ with $^*$), and the result is shown in Figure \ref{fig: Mixing_Length} after Gauss smoothing. 
Note that the vertical axis is $\lambda^* /Re_\tau^*$, and that the critical ratio $n$ in criterion \eqref{eq:diagnostic_function} is set to $2$. This is because at the outer edge of the boundary layer, $\partial u /\partial y$ is really small, leading to an ill-defined $\lambda^*$. 
Yet for the same reason, the magnitude of $\lambda^*$ in these regions is not important, as $\partial u/\partial y$ is negligibly small, and the integration in this region has little impact on the mean profile of $u^*$ or $\tau^*$. 

Twenty-five $\lambda^* $ profiles in five flow fields are presented in Figure \ref{fig: Mixing_Length}. Their growth trends and final stable values are closely aligned, showing obvious similarity in $\lambda^*$ across different flow fields. 
In flow fields with stronger pressure gradient, such as flow m18n, $\lambda^*$ in the outer region is less stable, but all the profiles end up at approximately $\lambda^*/Re_\tau^*= 0.08$, which agrees well with our current model \eqref{eq:lambdastar} at $y^* =Re_\tau^*$, as long as $Re_\tau ^* \gg a$ and $\ b\ll 1$ so that $\lambda^*(Re_\tau^*) \approx kb Re_\tau^*$, indicating that $kb\approx0.08$. This is also consistent with the optimal parameters in Table \ref{tab:Optimal_parameters}.

\begin{figure}[htbp]
    \centering
    \includegraphics[width=\textwidth]{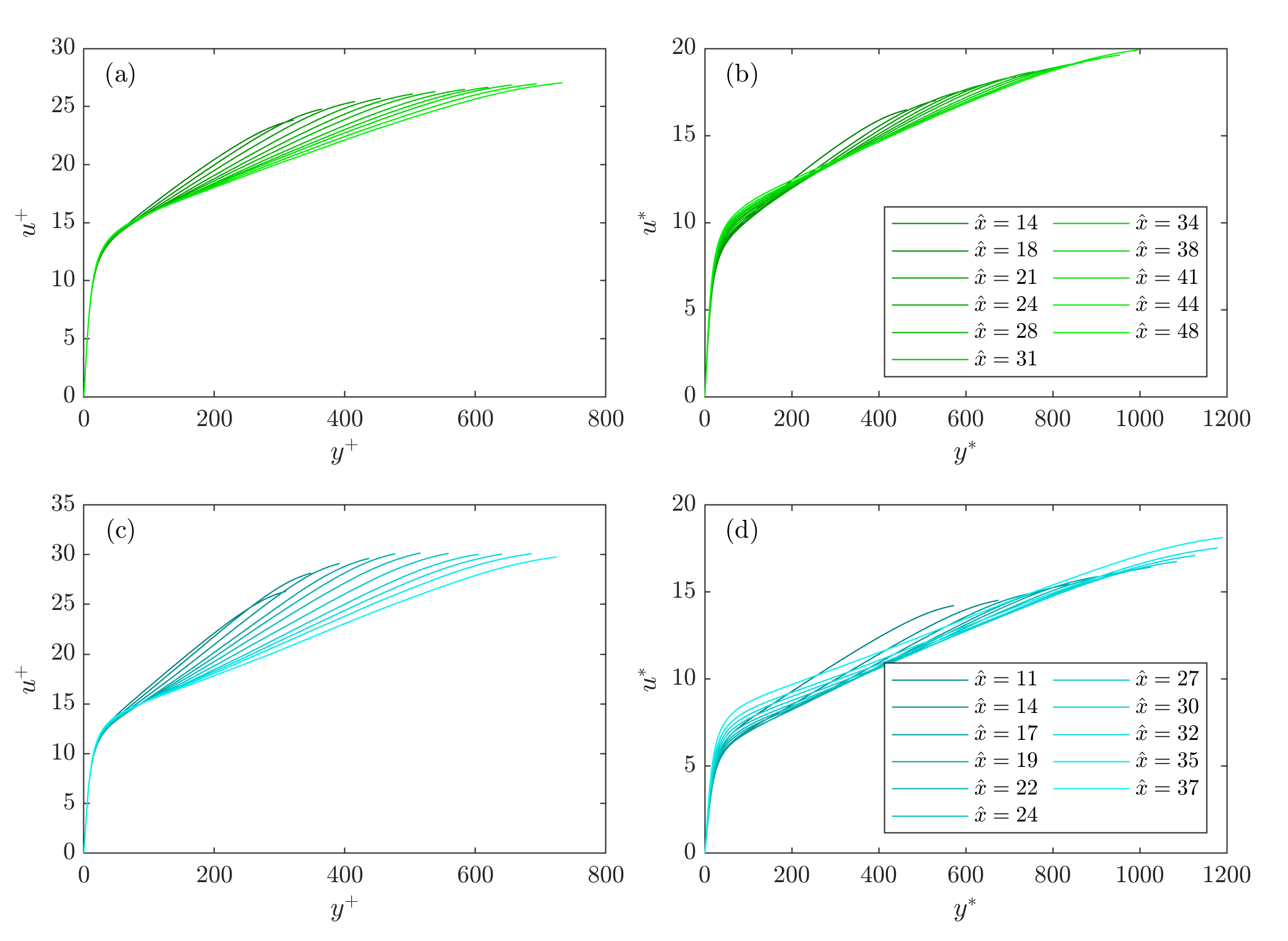}
    \caption{Profiles of $u^+$ and $u^*$ at different streamwise locations. (a),(b) denotes different profiles in flow b1n, and (c),(d) denotes different profiles in flow m16n. The data are obtained directly through LES \citep{Bobke_Vinuesa_Örlü_Schlatter_2017}. (a),(c) shows the profile of $u^+$ versus $y^+$; (b),(d) shows the profile of $u^*$ versus $y^*$. }
    \label{fig:Comparison of u star and u plus}
\end{figure}

To directly examine the streamwise self-similarity of mean velocity profiles within the same flow field, the profiles of $u^*$ with respect to $y^*$ at different streamwise locations are plotted in Figure \ref{fig:Comparison of u star and u plus}(b)(d). For comparison, we also plotted the profiles of $u^+$ with respect to $y^+$ in Figure \ref{fig:Comparison of u star and u plus}(a)(c). 
It can be clearly seen that, in Figure \ref{fig:Comparison of u star and u plus}(a)(c), although a consistent slope is observed in the inner region using traditional viscous coordinates \eqref{eq:friction_velocity_based_nondimensionalization}, velocity profiles in the outer region show poor agreement, where $\partial u^+/\partial y^+$ significantly decreases along streamwise direction. In contrast, as shown in Figure \ref{fig:Comparison of u star and u plus}(b)(d), the new scaling method \eqref{eq:u_star_based_nondimensionalization} overall makes the velocity profiles more consistent. In other words, the new scaling fix the streamwise self-similarity disrupted by PG effect.

\subsection{Verification of the Estimation-Correction Model}\label{subsec:Iteration Model}

In this section, it will be shown that the current estimation-correction model largely captures the shape of the profile and provides good predictive capability. 
To begin with, $\tau^*$ from equation \eqref{eq:tau1_y} can be decomposed into four parts, as shown in \eqref{eq:Understanding}, namely $\tau_{0}^*, \ \tau_{\beta}^*, \tau_{e}^*,\ \tau_{\gamma}^*$:
\begin{gather}
    \tau^*=\tau_{0}^*+\tau_{\beta }^*+\tau_{e}^*+\tau_{\gamma}^*,
    \label{eq:Understanding}\\
    \left\{
    \begin{aligned}
        \tau_{0}^* =&1-G(y^*),\\
        \tau_{\beta}^* =&\frac{\beta}{1+\beta}\frac{1}{\delta_1^*}\left(\theta^*(1-G(y^*))-\int_{y^*}^{Re_\tau^*}\left(1-\frac{\widetilde u^{*2}}{u_e^{*2}}\right)\mathrm d y^*\right),\\
        \tau_{e}^*=&\gamma\frac{\partial u_e^*}{u_e^*\partial \beta}\left(-\int_0^{y^*}\widetilde u^{*2}\mathrm dy^*+G(y^*)\int_0^{Re_\tau^*}\widetilde u^{*2}\mathrm dy^*\right),\\
        \tau^*_{\gamma}=&\gamma \left(
        \int_0^{y^*}\frac{\partial \widetilde u^{*2}}{\partial \beta}\mathrm d y^*
        -\widetilde u^*\int_0^{y^*}\frac{\partial \widetilde  u^*}{\partial \beta}\mathrm d y^* 
        -G(y^*)\left(
        \int_0^{Re_\tau^*} \frac{\partial \widetilde u^{*2}}{\partial \beta}
        \mathrm d y^*
        -u_e^*\int_0^{Re_\tau^*}\frac{\partial \widetilde u^*}{\partial \beta}\mathrm d y^*
        \right)
        \right).
    \end{aligned}
    \right.
    \nonumber
\end{gather}

The first part $\tau_{0}^*$ is the shear stress introduced by the natural increase in the friction Reynolds number, independent of pressure gradient.
It is also referred to as the Reynolds number effect, exactly the same form as equation \eqref{eq:ZPGTBL_tau}, which is the balanced ZPG TBL case. 
From equation \eqref{eq:tau1_y}, it's easy to verify that $G(0)=0$, and $G(Re_\tau^*)=1$, yielding $\tau_{0}^*(0)=1$, and $\tau_{0}^*(Re_\tau^*)=0$, which is consistent with ZPG boundary conditions.

The second part $\tau_{\beta}^*$ demonstrates the effect of equilibrium PG on the entire profile, which is independent of non-equilibrium term $\gamma$. The $(1+\beta)$ in the denominator is the result of scaling by $u_\beta^{2} = u_\tau^2 (1+\beta)$, and actually if scaled by traditional viscous coordinate, $\tau_{\beta}^+$ is directly proportional to $\beta$. At the boundary, $\tau_{\beta}^*(0)=-\beta/(1+\beta)$, and $\tau_{\beta}^*(Re_\tau^*)=0$. Together with $\tau_{0}^*$, we have $\tau_{0}^*(0)+\tau_{\beta}^*(0)=1/(1+\beta)$, which is consistent with wall conditions $\tau^*(0)=\tau/(\rho u_\beta^{2})=1/(1+\beta)$.

The third part $\tau_{e}^*$ and the fourth part $\tau_{\gamma}^*$ are both proportional to $\gamma$, which do not appear in the RANS equations, and these two parts are considered as the representation of first-order history effects. $\tau_{e}^*$ is also proportional to $\partial u_e^{*}/\partial \beta$, therefore representing the coupling effect between PG and the outer flow development, while $\tau_{\gamma}^*$ is related to the integral of $\partial u^*/\partial \beta$, which indicates the local non-equilibrium effect induced by the varying PG.
We have $\tau_{e}^*(0)=\tau_{e}^*(Re_\tau^*)=\tau_{\gamma}^*(0)=\tau_{\gamma}^*(Re_\tau^*)=0$, indicating that flow history only impacts regions far from the wall and far from the outer edge.

\begin{figure}[htbp]
    \centering
    \includegraphics[width=\textwidth]{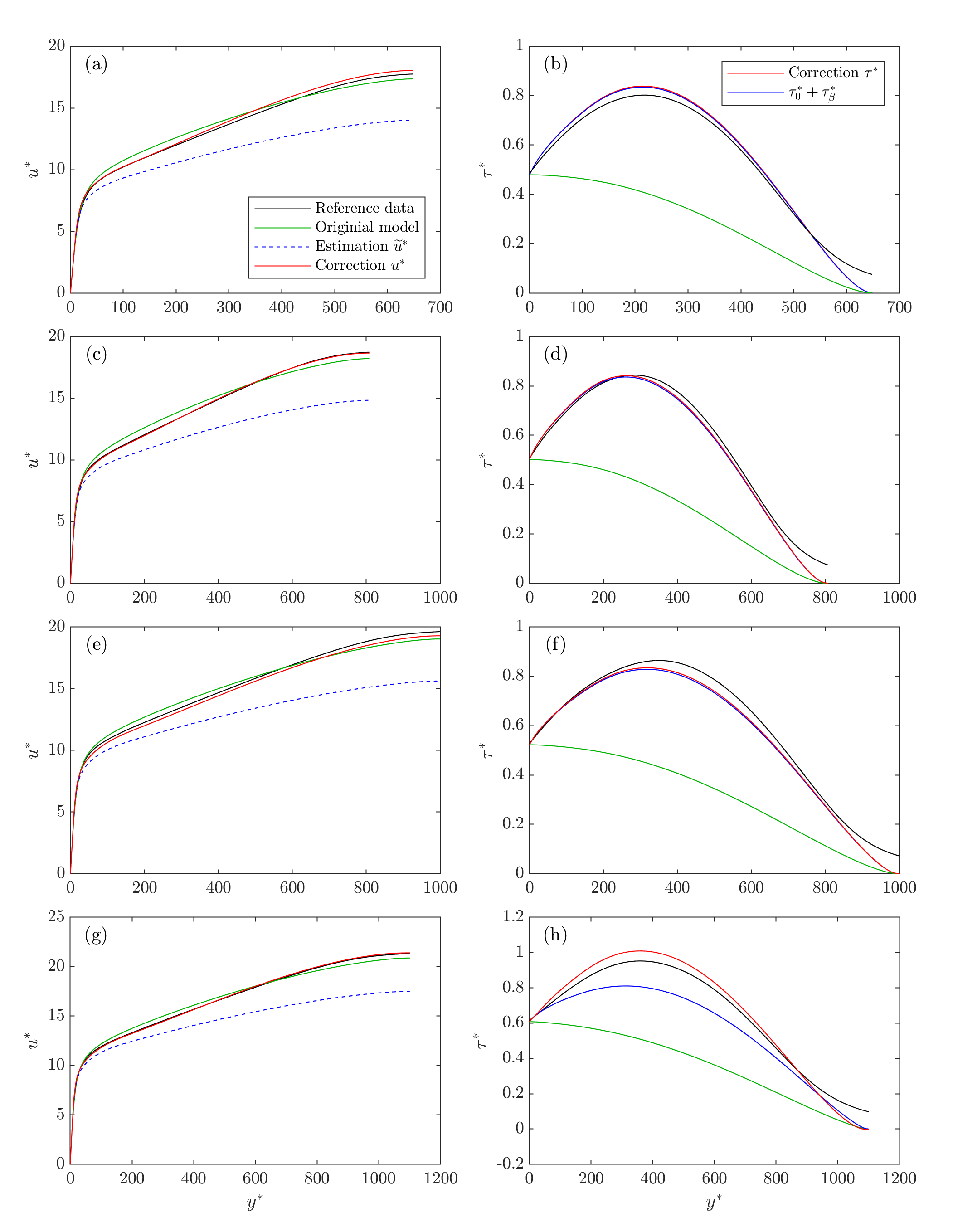}
    \caption{Velocity and total shear stress profiles in flow field b1n \citep{Bobke_Vinuesa_Örlü_Schlatter_2017}. (a)(b), (c)(d), (e)(f), (g)(h) show different streamwise locations respectively. (a)(c)(e)(g) are the profiles of mean velocity. The black solid line represents the reference LES data, the green solid line represents the original model by \cite{Subrahmanyam_Cantwell_Alonso_2022}, the blue dashed line represents the estimation $\widetilde u^*$ obtained from equation \eqref{eq:u1}, and the red solid line represents the correction $u^*$ obtained from equation \eqref{eq:u2(y)}. (b)(d)(f)(h) are the profiles of total shear stress. The black line represents the reference LES data, the green line represents the original model for ZPG TBLs by \cite{Subrahmanyam_Cantwell_Alonso_2022}, the red line represents the present model $\tau^*$ from \eqref{eq:tau1_y}, and the blue line represents its decomposition $\tau_{0}^*+\tau_{\beta}^*$, from equation \eqref{eq:Understanding}. The specific flow parameters are listed below: \\
    (a)(b) $\hat x=21, \beta = 1.1, \gamma = -1.4\times 10^{-5}, Re_\tau = 448$, 
    (c)(d) $\hat x=29, \beta = 1.0, \gamma = -1.2\times 10^{-5}, Re_\tau = 568$,\\
    (e)(f) $\hat x=41, \beta = 0.9, \gamma = -0.9\times 10^{-5}, Re_\tau = 716$,
    (g)(h) $\hat x=52, \beta = 0.6, \gamma = -14.7\times 10^{-5}, Re_\tau = 851$.
    }
    \label{fig:b1n}
\end{figure}

\begin{figure}[htbp]
    \centering
    \includegraphics[width=\textwidth]{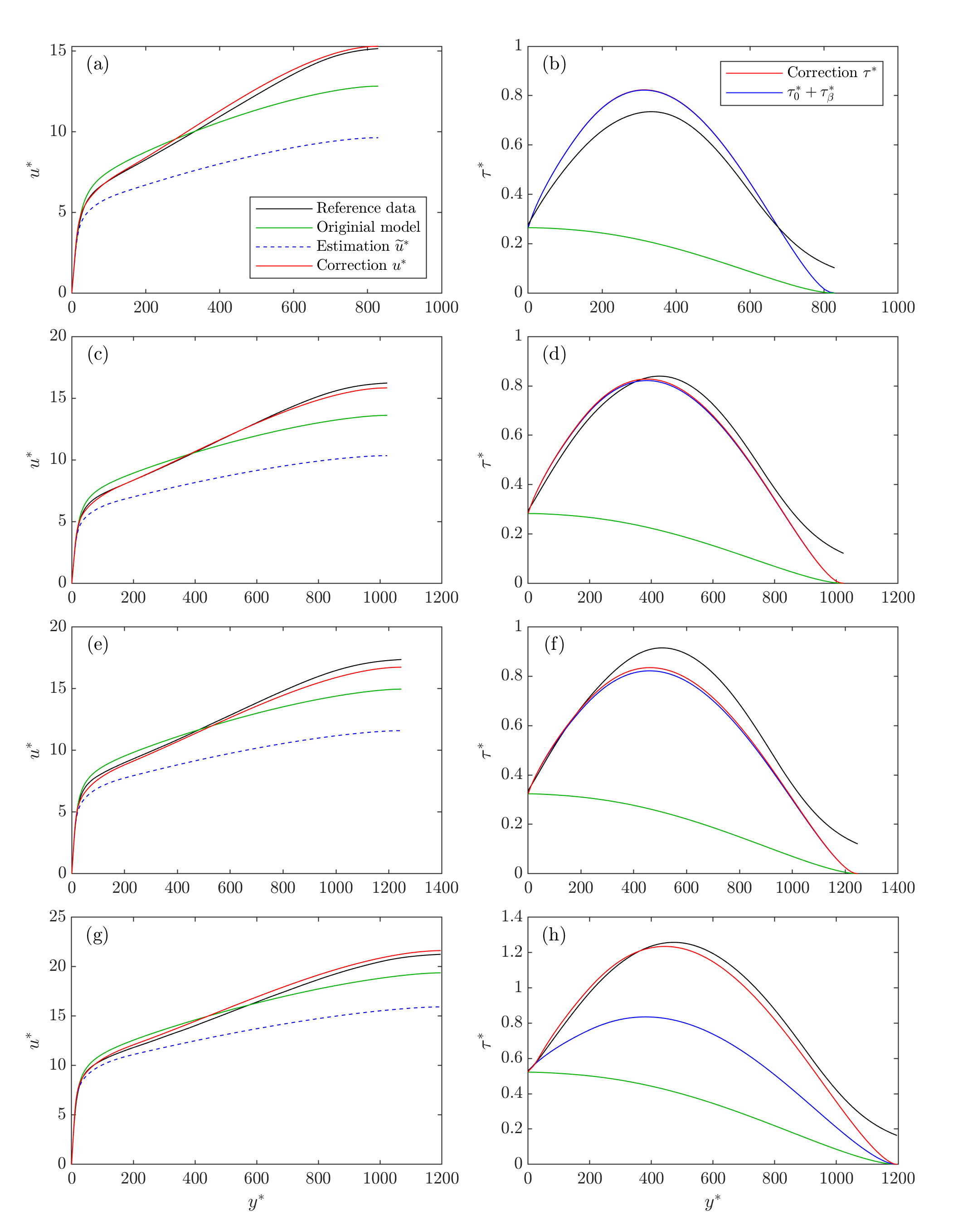}
    \caption{Velocity and total shear stress profiles in flow field m16n \citep{Bobke_Vinuesa_Örlü_Schlatter_2017}. (a)(b), (c)(d), (e)(f), (g)(h) show different streamwise locations respectively. The legend follows the same convention as in Figure \ref{fig:b1n}.
    The specific flow parameters are listed below: \\
    (a)(b) $\hat x=17, \beta = 2.8, \gamma = -1.1\times 10^{-5}, Re_\tau = 422$,
    (c)(d) $\hat x=23, \beta = 2.5, \gamma = -4.0\times 10^{-5}, Re_\tau = 539$,\\
    (e)(f) $\hat x=32, \beta = 2.1, \gamma = -4.8\times 10^{-5}, Re_\tau = 704$,
    (g)(h) $\hat x=41, \beta = 0.9, \gamma = -50.0\times 10^{-5}, Re_\tau = 861$.
    }
    \label{fig:m16n}
\end{figure}

\begin{figure}[htbp]
    \centering
    \includegraphics[width=\textwidth]{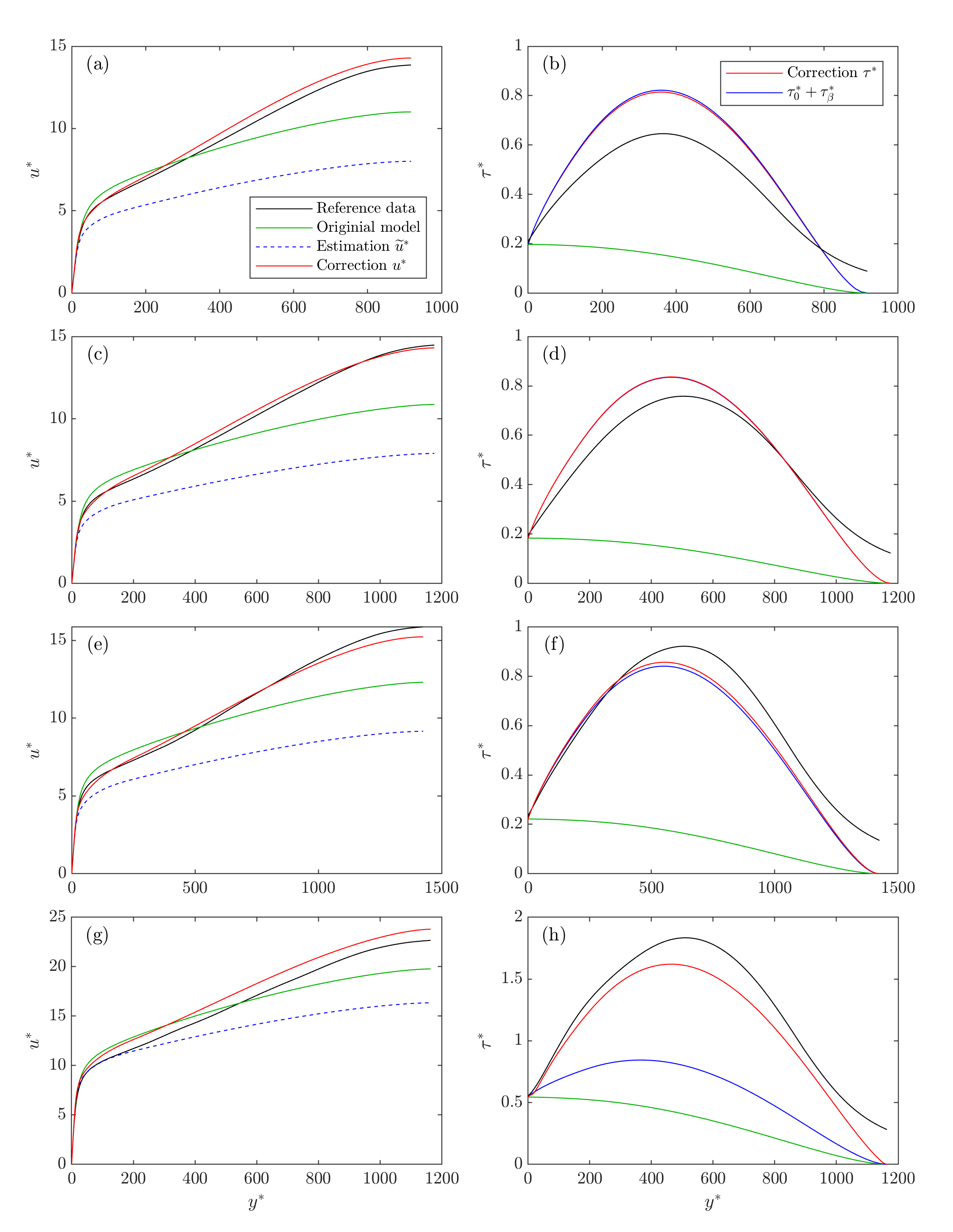}
    \caption{Velocity and total shear stress profiles in flow field m18n \citep{Bobke_Vinuesa_Örlü_Schlatter_2017}. (a)(b), (c)(d), (e)(f), (g)(h) show different streamwise locations respectively. The legend follows the same convention as in Figure \ref{fig:b1n}.
    The specific flow parameters are listed below: \\
    (a)(b) $\hat x=15, \beta = 4.1, \gamma = 14.6\times 10^{-5}, Re_\tau = 402$,
    (c)(d) $\hat x=21, \beta = 4.5, \gamma = -2.4\times 10^{-5}, Re_\tau = 503$,\\
    (e)(f) $\hat x=30, \beta = 3.5, \gamma = -15.9\times 10^{-5}, Re_\tau = 669$,
    (g)(h) $\hat x=38, \beta = 0.8, \gamma = -95.4\times 10^{-5}, Re_\tau = 858$.
    }
    \label{fig:m18n}
\end{figure}

\begin{figure}[htbp]
    \centering
    \includegraphics[width=\textwidth]{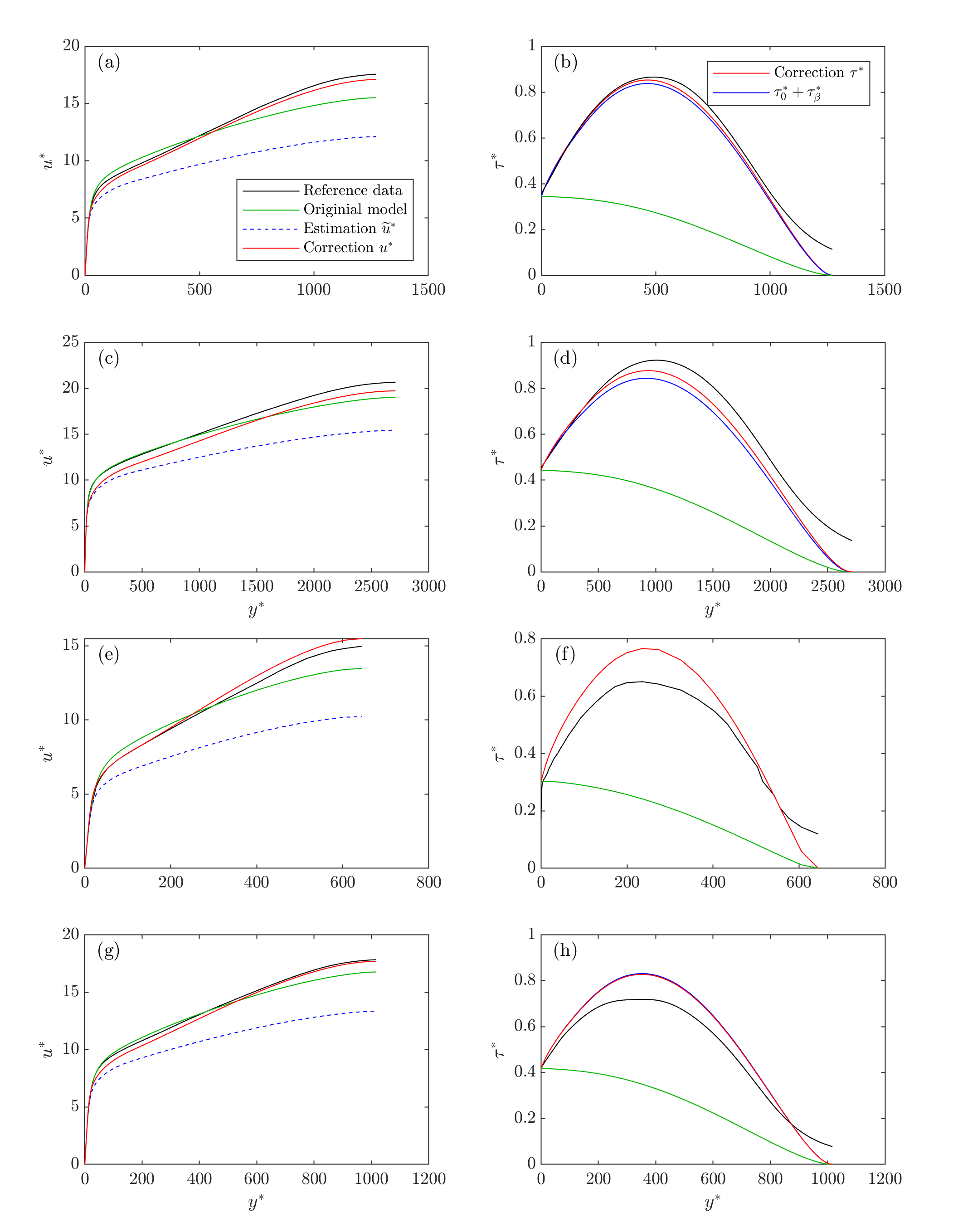}
    \caption{Velocity and total shear stress profiles in other flow fields. (a)(b), (c)(d), (e)(f), (g)(h) represent different flow fields respectively. The legend follows the same convention as in Figure \ref{fig:b1n}.
    The specific flow parameters are listed below:\\
    (a)(b) flow b2n \citep{Bobke_Vinuesa_Örlü_Schlatter_2017}, $\hat x=37, \beta = 1.9, \gamma = -5.5\times 10^{-5}, Re_\tau = 745$, \\ 
    (c)(d) flow b14 \citep{Pozuelo_Li_Schlatter_Vinuesa_2022}, $\hat x=46.5, \beta = 1.3, \gamma = -3.0\times 10^{-5}, Re_\tau = 1792$, \\
    (e)(f) flow LSM2017 \citep{Lee_2017}, $ \beta = 2.3, \gamma \sim0, Re_\tau = 364$,\\ 
    (g)(h) flow LSM2018 \citep{Yoon_Hwang_Sung_2018}, $\hat x=34, \beta = 1.4, \gamma = 1.0\times 10^{-5}, Re_\tau = 653$.
    }
    \label{fig:other verification}
\end{figure}

The validation results are shown in Figure \ref{fig:b1n}, \ref{fig:m16n}, \ref{fig:m18n}, \ref{fig:other verification}. 
Among them, Figure \ref{fig:b1n} shows the results of flow b1n, a rather mild APG flow, Figure \ref{fig:m16n} shows the results of flow m16n, a moderate APG flow, Figure \ref{fig:m18n} shows the results of flow m18n, a rather strong APG flow, and Figure \ref{fig:other verification} shows the results of other flow fields. 
In all four figures, the four subfigures on the left represent the validation results of mean velocity profiles, where the black line shows reference data obtained directly from DNS or LES, the green line shows Subrahmanyam's original model (\citeyear{Subrahmanyam_Cantwell_Alonso_2022}), the blue dashed line denotes the estimation-step result $\widetilde u^*$ derived from equation \eqref{eq:u1}, and the red line denotes the correction-step result $u^*$ derived from equation \eqref{eq:u2(y)}. Note that, the parameter $(k,a,m,b,n)$ of Subrahmanyam's model that we call here is the average of the parameter sets given in his original paper (Section 6, Table 5). 
The four subfigures on the right represent the validation results of total shear stress, where the black line still shows reference data from DNS or LES, the green line shows Subrahmanyam's original model (\citeyear{Subrahmanyam_Cantwell_Alonso_2022}) for ZPG TBLs, the red line denotes the correction-step result $\tau^*$ from \eqref{eq:tau1_y}, and the blue line denotes its decomposition $\tau_{0}^*+\tau_{\beta}^*$ from equation \eqref{eq:Understanding}, which is the part independent of $\gamma$. The discrepancy between the red line and the blue line is actually the first-order history effect that is proportional to $\gamma$ given by our estimation-correction model.

In Figure \ref{fig:b1n}(a)(c)(e)(g), it can be seen that the prediction of the correction-step result $u^*$ is highly accurate at all locations, generally superior to the original model, and far superior to the estimation-step result $\widetilde u^*$. 
Three models all behave well in the inner region. In the buffer region, the original model is larger, while $\widetilde u^*$ is smaller, and in the wake region, the original model and $\widetilde u^*$ are both smaller than LES result. This is because the initial estimation $\widetilde \tau^*$ in \eqref{eq:tau0}, which is obtained based on channel flow results, is a straight line, whereas in an APG TBL, real shear stress is a concave function that exceeds the initial $\widetilde \tau^*$ at all points except the boundaries, as shown in Figure \ref{fig:b1n}(b)(d)(f)(h). 
As a result, the original model and $\widetilde u^*$ which are based on $\widetilde \tau^*$ both underestimate. Meantime, in order to balance this outer-region error, based on the min-square-error criterion, the original model forcibly obtain the optimal parameters, resulting in the non-physical overestimation in the buffer region. In short, the advantage of our model $u^*$ is significant. However, the accuracy of $u^*$ on flow b1n is within expectation, as the tunable parameters $(k,a,m,b,n)$ of the our current model were trained based on this dataset.

In Figure \ref{fig:b1n}(b)(d)(f)(h), it can be seen that the correction-step prediction $\tau^*$ is also very close to LES results. In the inner region, at locations with smaller $|\gamma|$ where the boundary layer is in near-equilibrium state, such as Figure \ref{fig:b1n}(d)(f), $\tau^*$ agrees well with LES data, while in contrast, in Figure \ref{fig:b1n}(b)(h), where local non-equilibrium effect is rather significant, $\tau^*$ shows worse agreement. But, it's obvious that all profiles are markedly more accurate than the original model for ZPG TBLs, which completely fails to capture the first-increasing-then-decreasing trend of shear stress in APG TBLs. However, in the outer region, none of the profiles match the data well. One of the key reasons for this is that, although we have set the critical ratio $n = 5$ in criterion \eqref{eq:diagnostic_function} to ensure that the shear stress tends to  zero outside the boundary layer, it still remains around $0.1$ as shown by the black lines. However, in theoretical analysis, $\tau^*(Re_\tau^*)$ should accurately be zero. 

Furthermore, shown by the discrepancy between the red line and the blue line in Figure \ref{fig:b1n}(b)(d)(f)(h), the term $\tau_{e}^* + \tau_{\gamma}^*$ that indicates history effect, always adjusts $\tau^*$ in the correct direction, although it still fails to achieve a good match with the LES data. Upstream, the adjustment is always insufficient to work, since $\gamma$ is rather small, while downstream, it's sometimes too large. However, in Figure\ref{fig:b1n}(h), the effect of the streamwise derivative of pressure gradient on $\tau^*$ profile is still clearly visible.

The results in Figure \ref{fig:m16n} are generally consistent with those in Figure \ref{fig:b1n}. As shown in Figure \ref{fig:m16n}(a)(c)(e)(g), the correction-step result $u^*$ maintains high accuracy across all locations, and its improvement over the estimation $\widetilde u^*$ and the original model is even more evident.
In Figure \ref{fig:m16n}(b)(d)(f)(h), although the final result $\tau^*$ exhibits some numerical deviation, it still captures the overall trend of the LES data well. It is worth noting that flow field m16n is part of the validation set rather than the training set, so that all well-matched profiles indicate that our model exhibits high accuracy and generalizability.

Figure \ref{fig:m18n} still presents almost the same results as Figure \ref{fig:b1n} and Figure \ref{fig:m16n}. In Figure \ref{fig:m18n}(a)(c)(e)(g), the prediction of $u^*$ is still accurate at all locations, far better than $\widetilde u^*$ and the original model. In Figure \ref{fig:m18n}(b)(d)(f)(h), $\tau^*$ still captures the general shape and trend of the LES data, despite exhibiting non-negligible deviations in magnitude. 
In general, for flow m18n, the prediction deviates more from LES data compared to flow b1n or m16n. This can be attributed partially to the stronger APG effect, which induces more significant local non-equilibrium effects, and requires higher order modifications. However, as a validation set, the observed error of $u^*$ in Figure \ref{fig:m18n}(e)(g) remains within an acceptable range. 

In Figure \ref{fig:m18n}(b)(d)(f)(h), the history effect terms $\tau_{e}^*$ and $\tau_{\gamma}^*$ still adjust $\tau^*$ in the correct direction, and especially, in Figure \ref{fig:m18n}(h) where $|\gamma|$ is considerably large, history effect plays a significantly vital role in the total shear stress profile. However, the current stress profile predictions still leave room for improvement. One possible reason for the overprediction in Figure \ref{fig:m18n}(b)(d) may be the relative low Reynolds number($Re_\tau=402,503$), which leads to incomplete turbulence development and, consequently, an overestimation of the Reynolds stress $\langle u'v'\rangle$. Another possible remedy may be the consideration of higher-order history effects, which will be further discussed in Section \ref{sec:Conclusion}.

To avoid excessive discussion, the validation results for other flow fields are not presented in such detail. Instead, four selected profiles are shown in Figure \ref{fig:other verification}. The validation results in these four flow fields are almost identical to those in b1n, m16n or m18n. $u^*$ remains in good agreement with the LES or DNS results, and 
$\tau^*$ also successfully captures the general shape, including the location of the stress peak. However, noticeable discrepancies with the numerical results still remain, and the considerable overprediction in Figure \ref{fig:other verification}(f)(h) may still result from the relatively low Reynolds number.

\begin{figure}[htbp]
    \centering
    \includegraphics[width=\textwidth]{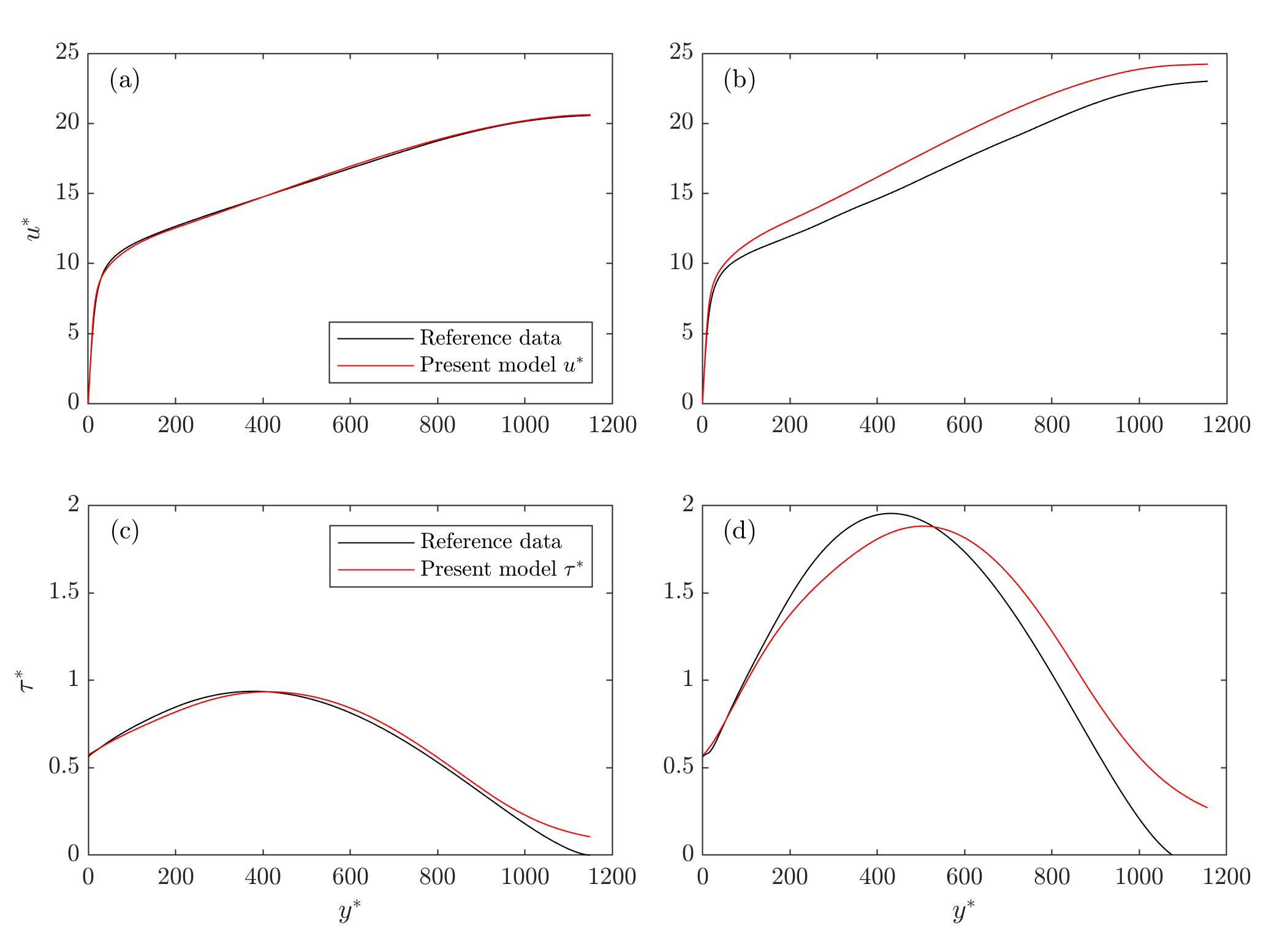}
    \caption{
    Two profiles extracted from flow field m13n and m18n \citep{Bobke_Vinuesa_Örlü_Schlatter_2017}, respectively, with the same $\beta$ and $Re_\tau$, but different $\gamma$. Subfigures (a) and (b) show the mean velocity profiles, and  (c) and (d) show the total shear stress profiles. The black solid lines represent LES results, and the red solid lines represent corrected prediction from the current model.\\
    (a)(c): flow m13n, $\hat{x} = 48$, $\beta = 0.77$, $\gamma = -11.1 \times 10^{-5}$, $Re_\tau = 862$, \\
    (b)(d): flow m18n, $\hat{x} = 37.8$, $\beta = 0.77$, $\gamma = -97.6 \times 10^{-5}$, $Re_\tau = 861$.
    }
    \label{fig:Same comparison}
\end{figure}

To intuitively demonstrate the role of history effect, we extract one profile each from the flow field m13n and m18n. These two profiles share the same $Re_\tau$ and $\beta$, but differ significantly in $\gamma$, with a ratio of approximately 9. The streamwise locations of the selected profiles are indicated by the black square markers in Figure \ref{fig: Overall_profile}(a). The corresponding computational results are shown in Figure \ref{fig:Same comparison}. The two subfigures on the left, (a) and (c), correspond to the profile from m13n, while the two on the right, (b) and (d), correspond to the profile from m18n. The top row, subfigures (a) and (b), present the mean velocity profiles, while the bottom row, (c) and (d), present the total shear stress profiles. In all subfigures, the black solid lines represent the LES results, and the red solid lines represent the corrected predictions from our model.

By comparing Figure \ref{fig:Same comparison}(a) with (b), or (c) with (d), it is evident that both the velocity and shear stress distributions differ significantly between the two profiles. In m18n, the velocity profile is lower near the wall and higher in the outer region, while the shear stress profile is noticeably higher across the entire domain. Encouragingly, our model, through the correction involving $\gamma$, successfully captures the differences between the two profiles.

\begin{figure}[htbp]
    \centering
    \includegraphics[width=\textwidth]{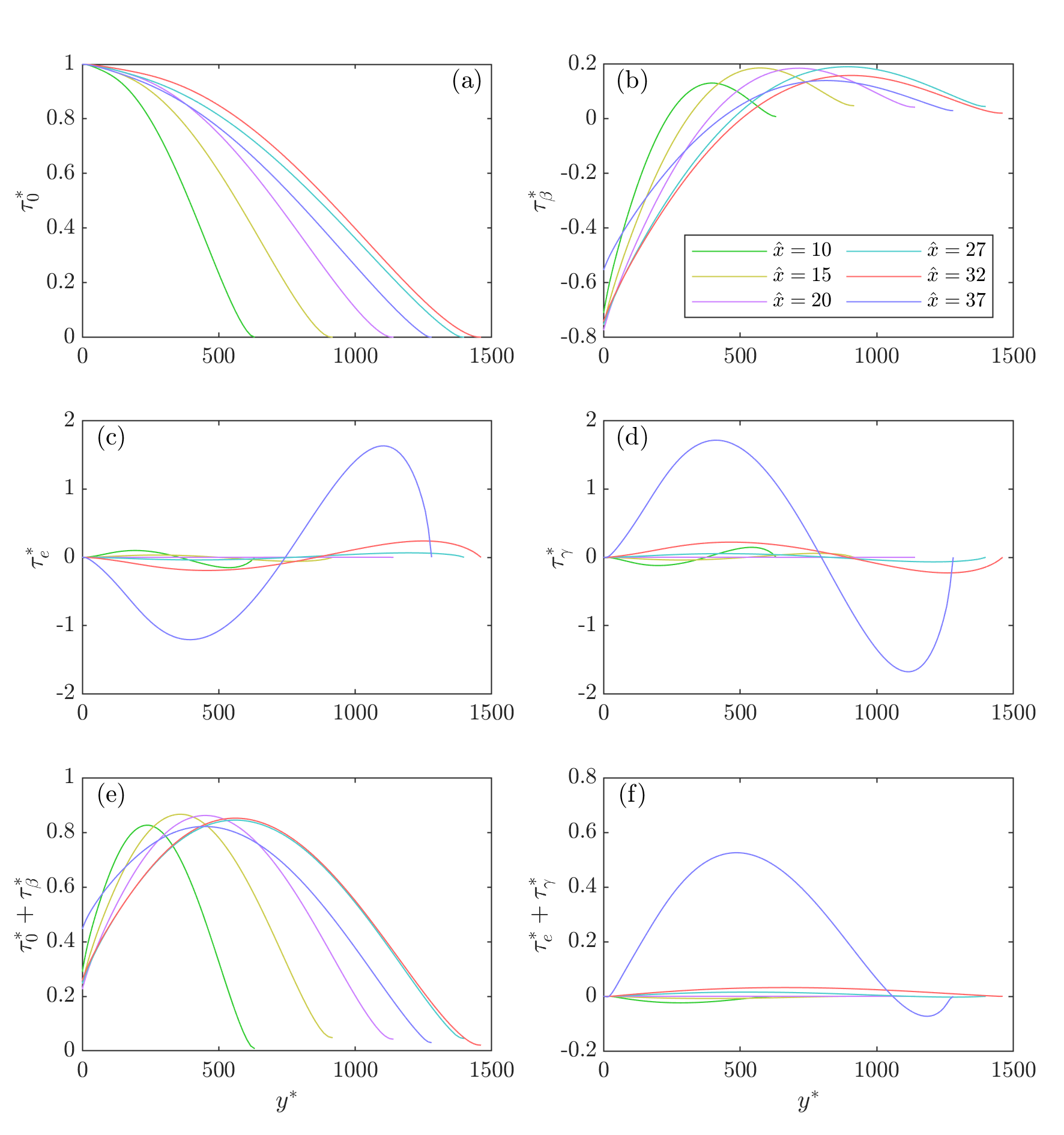}
    \caption{Profiles of the components of the correction-step result $\tau^*$, calculated at different streamwise locations of flow field m18n \citep{Bobke_Vinuesa_Örlü_Schlatter_2017}. (a)(b)(c)(d) shows the contribution of $\tau_{0}^*,\tau_{\beta}^*,\tau_{e}^*$, and $\tau_{\gamma}^*$, respectively, (e) shows $\tau_{0}^* +\tau_{\beta}^*$, and (f) shows $\tau_{e}^*+\tau_{\gamma}^*$.}
    \label{fig:Decomposition}
\end{figure}

Finally, a detailed analysis was performed on the contributions of each decomposed term of $\tau^*$ in equation \eqref{eq:Understanding}, based on the data calculated at six different streamwise locations of flow m18n. The corresponding results are shown in Figure \ref{fig:Decomposition}.

Figure \ref{fig:Decomposition}(a) shows the profile of $\tau_{0}^*$, which is the part completely independent of pressure gradient. Its profile is similar to the ZPG TBL case, or even the straight-line profile of channel flow case. This explains why the channel flow shear stress profile can be used directly in ZPG TBLs, and can be used as the initial estimation in our model. All the six profiles start from $\tau_{0}^*(0)=1$ at the wall, and monotonically decrease to $\tau_{0}^*(Re_\tau^*)=0$ at the outer edge. The profiles are highly similar, except for the difference in friction Reynolds number which determines the outer edge position.

Figure \ref{fig:Decomposition}(b) shows the profile of $\tau_{\beta}^*$, which is the part directly proportional to $\beta$ and independent of non-equilibrium effect $\gamma$. All the six profiles start from $\tau_{\beta}^*(0)=-\beta/(1+\beta)$, reach a peak of about $0.1\sim0.2$, and finally drop to zero at the outer edge. The peak value gradually decreases along the streamwise direction. Despite the difference in $\beta$, the six profiles still exhibit strong similarity.

Figure \ref{fig:Decomposition}(c),(d) show the profile of $\tau_{e}^*$ and $\tau_{\gamma}^*$ respectively, which are the parts directly proportional to $\gamma$. These two terms are almost equal in magnitude but opposite in sign, achieving a delicate balance between the interaction of PG with outer flow development, and the local pressure non-equilibrium effect. All the profiles vary from zero back to zero, undergoing a transition from positive to negative, or the opposite, depending on the sign of $\gamma$. 

Figure \ref{fig:Decomposition}(e),(f) show the profile of $\tau_{0}^* +\tau_{\beta}^*$, which is the part that is independent of $\gamma$, and the profile of $\tau_{e}^*+\tau_{\gamma}^*$, which is the part that is proportional to $\gamma$ and shows first-order history effect. A comparison between the two subfigures reveals that the non-history effect component of the total stress is relatively consistent across all profiles, while the history-effect term varies sensitively with the magnitude of $\gamma$. The history effect peaks at approximately 0.5, contributing about half of the total stress.

\section{Discussion and Conclusion}\label{sec:Conclusion}

In this study, a general modeling framework for APG TBLs is developed, which, for the first time, explicitly incorporates history effect through local flow parameters. A modified set of non-dimensional scales successfully recovers the streamwise self-similarity disrupted by the pressure gradient to some extent. The streamwise derivative of pressure gradient, $\gamma$, is naturally introduced in the estimation-correction method between $u$ and $\tau$, which can represent the accumulated upstream history effect, significantly enhancing the predictive capability for both mean velocity and total shear stress profiles.

The correction-step result $\tau^*$ is decomposed into four parts, corresponding respectively to Reynolds number effects, equilibrium pressure gradient effects, the coupling between free-stream velocity and pressure gradient, and local non-equilibrium pressure gradient effects. The latter two are considered first-order history effects, which typically offset each other, but in downstream regions their combined effect becomes significant, accounting for nearly half of the total shear stress.

The model is validated against eight DNS or LES datasets, covering a wide range of Reynolds numbers and pressure gradients. Across all cases, the model maintains high predictive accuracy, outperforms previous methods, and clearly demonstrates the impact of history effect on the boundary layer profiles.

In fact, the estimation-correction method can be further repeated, allowing for introducing higher order streamwise derivatives of PG into the profiles of velocity or shear stress. 
After closure via the mixing-length model, the original RANS equations contain only two unknowns, $u$ and $v$, while in the iterative estimation-correction method, we may iteratively solve for $u^*$ and $\tau^*$, which also constitute a system with two unknowns. This guarantees that this iterative estimation-correction model can make full use of the information given by the governing equations.
In each round of iteration, first, $u^{*}_{n+1}$ is obtained given $\tau^{*}_n$. This process can be completed directly through equation \eqref{eq:u(y)} without adding computational complexity. 
Then, just as in the current model, by converting the streamwise derivatives into the derivatives of other flow parameters, and then sequentially expanding the terms in RANS equations and the continuity equation, we can solve for $\tau_{n+1}^*$ according to $u^*$ as an explicit function. The iteration can be done systematically, introducing higher-order correction, and the terms can be regarded as higher-order history effects.
However, complexity grows rapidly as further expansion is conducted.

There also remain several limitations. Most notably, the predictions of the shear stress profiles are still not sufficiently accurate.
One of the issues is that the predicted shear stress remains nonzero at the outer edge of the boundary layer, which contradicts the theoretical assumptions. Future work may address this by redefining the boundary layer thickness $\delta$ or manually adding a correction term at the edge.
Furthermore, it is also possible to apply the iterative method discussed earlier, which may yield more precise results.

Although the present study focuses exclusively on APG TBLs, the theoretical derivation does not impose a strict constraint requiring $\beta > 0$. Therefore, theoretically, the model may be extended to FPG TBLs as well. However, the velocity scale in the model is defined as $u_\beta = u_\tau \sqrt{1 + \beta}$, which implies that $\beta$ must be greater than $-1$, severely limiting the applicability of the model to FPG cases. Furthermore, due to the turbulence-suppressing effect of FPG, current studies and datasets for FPG TBLs remain limited.

Additionally, although the model has been tested for cases with $\beta$ up to 4.53, it is not applicable to profiles near boundary layer separation, which is the area of great practical interest. One critical issue is that near separation, the wall friction velocity $u_\tau$ approaches zero, rendering all wall-based scalings invalid. 

In summary, our model accurately predicts the mean velocity and total shear stress profiles in APG TBLs and explicitly incorporates history effect through local parameters. This not only deepens the understanding of the underlying physics of APG TBLs but also offers new ideas and insights for improving turbulence modeling and closure strategies for the RANS equations. Furthermore, the present work may provide a theoretical foundation for the development of wall models in LES under complex boundary conditions.

\section*{Acknowledgement.}
The authors would like to express their appreciation to the program of Open Research for Innovation Challenges (ORIC), supported by Tsien Excellence in Engineering Program (TEEP), Tsinghua University.

\section*{Funding.}
This work is supported by the National Natural Science Foundation of China (under grant nos.
12272205 and 12388101).

\section*{Declaration of Interests.}
The authors report no conflict of interest.

\clearpage
\bibliographystyle{jfm}
\bibliography{reference}

\end{document}